\documentclass[fleqn,usenatbib]{mnras}
\usepackage{newtxtext,newtxmath}
\usepackage{booktabs}

\usepackage[T1]{fontenc}
\usepackage{ae,aecompl}

\usepackage{graphicx}	
\usepackage{amsmath}	
\usepackage{amssymb}	
\usepackage{titlesec}
\usepackage{epstopdf}
\usepackage{pdflscape}	
\usepackage[labelformat=parens,labelsep=none,subrefformat=parens]{subcaption}
\usepackage{color}
\usepackage{cleveref}
\usepackage{float}

\newcommand{\hi}{\ion{H}{i}}	
\newcommand{\hii}{\ion{H}{ii}}	
\newcommand{\halpha}{\ion{H}{$\alpha$}}

\newcommand{\bub}{\text{GSH 006$-$15$+$7}}
\newcommand{\kms}{\text{km\,s$^{-1}$}}
\newcommand{\radms}{\text{rad\,m$^{-2}$}}

\title[Ghost of a Shell: Magnetic Fields of Galactic Supershell GSH 006$-$15$+$7]{Ghost of a Shell: Magnetic Fields of Galactic Supershell \bub}

\author[A. J. M. Thomson et al.]{Alec J. M. Thomson,$^{1}$\thanks{E-mail: alec.thomson@anu.edu.au}
N. M. McClure-Griffiths,$^{1}$
Christoph Federrath,$^{1}$
\newauthor John M. Dickey,$^{2}$
Ettore Carretti,$^{3}$ B. M. Gaensler,$^{4}$ Marijke Haverkorn,$^{5}$
\newauthor
M. J. Kesteven,$^{6}$
Lister Staveley-Smith$^{7}$
\\
$^{1}$Research School of Astronomy and Astrophysics, Australian National University, Canberra, ACT 2611, Australia\\
$^{2}$School of Mathematics and Physics, University of Tasmania, Hobart TAS 7001, Australia\\
$^{3}$INAF, Osservatorio Astronomico di Cagliari, Via della Scienza 5, 09047 Selargius (CA), Italy\\
$^{4}$Dunlap Institute for Astronomy and Astrophysics, University of Toronto, 50 St. George Street, Toronto ON, M5S 3G4, Canada \\
$^{5}$Department of Astrophysics/IMAPP, Radboud University, P.O. Box 9010,
6500 GL Nijmegen, The Netherlands\\
$^{6}$CSIRO Astronomy and Space Science, PO Box 76, Epping, NSW 1710, Australia\\
$^{7}$International Centre for Radio Astronomy Research (ICRAR), The University of Western Australia, 35 Stirling Hwy, Crawley, \\
WA 6009, Australia\\
}

\date{Accepted 2018 July 9. Received 2018 July 9; in original form 2018 February 5}

\pubyear{2018}

\begin{document}
\label{firstpage}
\pagerange{\pageref{firstpage}--\pageref{lastpage}}
\maketitle

\begin{abstract}
We identify a counterpart to a Galactic supershell in diffuse radio polarisation, and use this to determine the magnetic fields associated with this object. \bub\ has perturbed the polarised emission at 2.3\,GHz, as observed in the S-band Polarisation All Sky Survey (S-PASS), acting as a Faraday screen. We model the Faraday rotation over the shell, and produce a map of Faraday depth over the area across it. Such models require information about the polarised emission behind the screen, which we obtain from the Wilkinson Microwave Anisotropy Probe (WMAP), scaled from 23\,GHz to 2.3\,GHz, to estimate the synchrotron background behind \bub. Using the modelled Faraday thickness we determine the magnitude and the plane-of-the-sky structure of the line-of-sight magnetic field in the shell. We find a peak line-of-sight field strength of $|B_\parallel|_\text{peak} = 2.0\substack{+0.01 \\ -0.7}\,\mu$G. Our measurement probes weak magnetic fields in a low-density regime (number densities of $\sim0.4\,$cm$^{-3}$) of the ISM, thus providing crucial information about the magnetic fields in the partially-ionised phase.
\end{abstract}





\section{Introduction}
\hi\ shells, bubbles, supershells, and superbubbles are large structures in the interstellar medium (ISM) blown out by hot OB star clusters and supernovae. Both a supernova and the winds from a massive star in its main sequence lifetime will each inject around $10^{51}$-$10^{53}$\,ergs into the ISM. These winds and shocks ionise what will become the cavity of the shell, and sweep out the neutral material \citep{McClure-Griffiths2002}. It is now understood that these objects are strongly influenced by magnetic fields in their formation \citep{Tomisaka1990,Tomisaka1998,Ferriere1991,Slavin1992,Ntormousi2017,Gao2015,Stil2009}. Magnetic fields both oppose the expansion of the shell from the exterior, and prevent the collapse of the swept up shell walls \citep{Ferriere2001}. \hi\ shells have been discovered throughout our Galaxy \citep{Hu1981a,Koo1992,Maciejewski1996,Uyaniker1999,McClure-Griffiths2000,McClure-Griffiths2002,Mcclure-Griffiths2006,Pidopryhora2007,Heiles1979}, as well as external galaxies. These objects play a large role in determining the dynamics, evolution, and overall structure of the ISM \citep{McClure-Griffiths2002}. 

Supershells and superbubbles are the largest classification of \hi\ shells, with radii between $10^2$ and $10^3$\,pc \citep{Heiles1979}. Such objects occupy an intermediate size-scale within the ISM; a scale at which the role of magnetic fields in the magneto-ionic medium (MIM) is not well understood. Supershells are most commonly found in \hi\ surveys, and appear as cavities in the neutral hydrogen \citep[e.g.][]{McClure-Griffiths2002}; but they can also have multi-wavelength properties. Supershells can have associated emission in \halpha, soft X-rays, far ultra-violet, polarised radio continuum, and 100\,$\mu$m \citep{Heiles1979,Moss2012,Heiles1999,McClureGriffiths2001,Heiles1984,McClure-Griffiths2002,Ehlerova2005,Ehlerova2013,Suad2014,Jo2011,Reynolds1998, Boumis2001,Jo2015}. It is thought that once bubbles expand far enough to break out of the gas of the Galactic plane, the shell breaks open into a Galactic chimney, allowing the flow of hot gas into the halo \citep{Norman1989}. This transition from bubble to chimney is slowed by the presence of magnetic fields, which tend to confine the expanding shell in the disc \citep{Tomisaka1998}.

A powerful method for probing the magnetic fields of the ISM is the study of Faraday rotation. This phenomenon describes how the polarisation angle (PA) of a linearly polarised wave will rotate as the wave propagates through a MIM. Faraday rotation is measured from the linear Stokes parameters, $Q$ and $U$. Following \citet{Burn1966} and \citet{Brentjens2005}, these can be parametrised as the complex polarisation, $\mathcal{P}$:
\begin{equation}
	\mathcal{P} = Q + iU = \text{PI} e^{2i\text{PA}}
\end{equation}
From this, the polarised intensity (PI) and the polarisation angle can therefore be defined as:
\begin{equation}
\begin{aligned}
	&\text{PI} = \sqrt{Q^2+U^2}\\
	&\text{PA} = \frac{1}{2}\arctan{\left(\frac{U}{Q}\right)}
\end{aligned}
\end{equation}
The amount of Faraday rotation at a wavelength $\lambda$ is described by the Faraday depth ($\phi$) times $\lambda^2$:
\begin{equation}
\phi(L)\equiv0.812\int_{0}^{L}n_e B_\parallel dr\text{\,\radms}
\label{eqn:faradaydepth}
\end{equation}
Where $B_\parallel$ is the line-of-sight magnetic field in $\mu$G, $n_e$ is the electron density in $\text{cm}^{-3}$, and $r$ is the distance along the line of sight integrated through the Faraday rotating medium with length $L$ in pc. The sign of $B_\parallel$ is taken to be positive when the field is aligned towards the observer and vice-versa. In the simplest Faraday rotation case, where all the emission is in the background and all the Faraday rotation occurs in the foreground, the Faraday depth is given by the rotation measure (RM):
\begin{equation}
\text{RM} = \frac{\Delta \text{PA} }{\Delta( \lambda^2) }
\end{equation}
This also assumes no depolarisation occurs along the line of sight. A method of obtaining Faraday depth in more complex scenarios is described in \citet{Brentjens2005}.

Despite the important role magnetic fields play in the Galactic ISM, several mysteries remain unresolved. Firstly, as summarised by \citet{Han2017}, the magnetic fields of a number of extended diffuse objects have been studied. However, obtaining the complete scale and structure of the magnetic fields associated with these objects is difficult. This arises as a result of both the methods used to measure these fields, and line-of-sight confusion from other magneto-ionic objects. Studies of RMs from both extragalactic point sources and pulsars can suffer from line-of-sight field reversals and confusion, and intrinsic Faraday rotation. These line-of-sight effects are a particular concern towards the Galactic plane. In this sense, diffuse polarisation studies have a particular advantage in revealing large-scale, extended structures.

\bub\ is a recently discovered Galactic supershell located near the Galactic plane \citep{Moss2012}. This supershell was discovered in \hi\ observations with a central velocity around $v_\text{LSR}\approx7\,$\kms, and subtends about $25\deg$ on the sky. \citet{Moss2012} constrain the age and distance to this object at $15\pm5\,$Myr and $1.5\pm0.5$\,kpc, respectively; this gives the shell an approximate diameter of $670 \pm 220\,$pc, making it one of the largest discovered \hi\ shells near the Sun. From their analysis, \citet{Moss2012} find that \bub\ is likely in a break-out phase between a supershell and a chimney structure.

In this paper we present the counterpart to \bub\ in diffuse polarised radio emission data at 2.3\,GHz. The shell appears as a `shadow' in polarised emission, and shows evidence of Faraday rotation of background synchrotron radiation. The data we use in this analysis is described in Section~\ref{sec:data}. In Section~\ref{sec:analysis} we discuss this morphological association found in diffuse polarisation at 2.3\,GHz. We continue to use these polarisation data to model the Faraday rotation through \bub\ as a Faraday screen, following \citet{Sun2007a} and \citet{Gao2015}. From this, in Section~\ref{sec:results} we constrain both the magnitude and the plane-of-sky structure of the line-of-sight magnetic fields associated with \bub. We use this information to estimate the thermal and magnetic pressures within the Galactic supershell. This dynamical information, as well as the magnetic field strengths themselves, will aid in the future modelling of supershells and the ISM, as well as in our overall understanding of ISM magnetohydrodynamics. Our conclusions are given in Section~\ref{sec:summary}.

\section{Data}\label{sec:data}

\subsection{Diffuse H\,{\scriptsize \textbf{I}} Emission}
\subsubsection{Galactic All-Sky Survey}
We make use of \hi\ data from the third release of the Parkes Galactic All-Sky Survey \citep[GASS,][]{McClure-Griffiths2009,Kalberla2009}, now incorporated in HI4PI \citep{BenBekhti2016}. GASS is a fully-sampled survey of \hi\ emission over the entire sky south of declination zero with an angular resolution of 14.4 arcmin. The survey was conducted with the Parkes radio telescope using the 13-beam multibeam receiver. The data cover the velocity range $-468 \leq v_{LSR} \leq 468~{\rm km~s^{-1}}$ with a velocity resolution of 1\,\kms and a typical rms noise of 57\,mK. In the third data release of GASS used here, the calibration and stray radiation correction were refined by \citet{Kalberla2015}. We use the GASS data to extract velocity separated \hi\ emission of \bub.

\begin{figure}
	\includegraphics[width=\columnwidth]{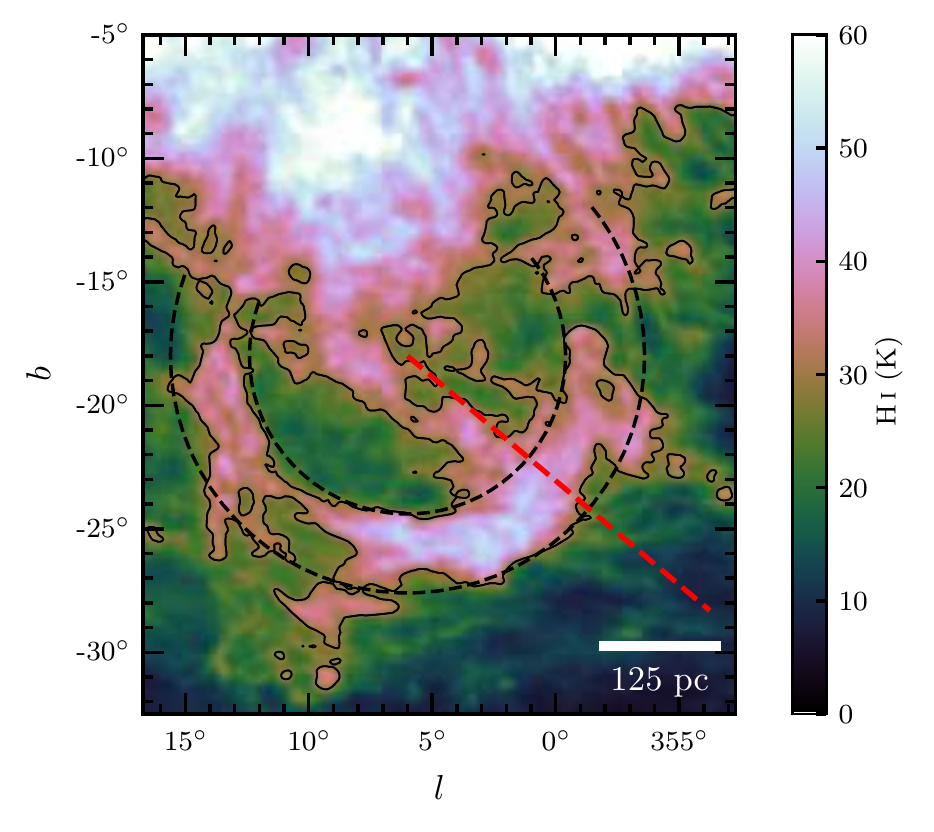}
	\caption{GASS -- Brightness temperature of the \hi\ 21\,cm line in the region of the \bub\ shell at $v_{LSR}=6.6$\,\kms. All maps are given in Galactic longitude and latitude ($l,b$). Contour given at $30\,$K, which well outlines the \hi\ emission associated with the shell. The black dashed lines give the approximate inner and outer bounds of the shell; approximately matching the $30\,$K contour. The scalebar gives the approximate size scale assuming a distance of 1.5\,kpc. The red dashed line in corresponds to the profile given in Figure~\ref{fig:profiles}. This profile was chosen to match the one given in Figure~\ref{fig:SPASSPIGASS}.}
	\label{fig:GASS}
\end{figure}

\subsection{Radio Continuum Polarisation}

\subsubsection{S-band Polarisation All Sky Survey}
The S-band Polarisation All Sky Survey (S-PASS, \citealt{Carretti2013}, Carretti et al. in preparation) was completed in 2010, and provides a highly sensitive polarisation (Stokes $Q$ and $U$) map of the Southern sky at 2.3\,GHz. The survey was conducted using the Parkes 64\,m Telescope with its `Galileo' receiver and covers the Southern sky at declinations $\delta<-1\deg$. This receiver operates in S-Band (13\,cm) and is sensitive to circularly polarised radiation; allowing for the linear Stokes parameters, $Q$ and $U$, to be measured. Table~\ref{tab:spassparams} lists the observational parameters for S-PASS. Initial morphological analysis was conducted by \citet{Carretti2013}, and here we provide additional morphological descriptions; specifically we find a morphological correlation between the structure of \bub\ in \hi\ and the structure found in Stokes $Q$ and $U$ from S-PASS. 

S-PASS supplies a number of significant improvements over previous polarisation surveys. Observations of polarised emission at 2.3\,GHz are inherently less prone to depolarisation effects \citep{Burn1966,Sokoloff1998} with respect to lower frequencies. This higher observing frequency allows for greater angular resolution, with S-PASS presenting a gridded angular resolution (full width at half maximum $\text{FWHM}$) of 10.75'. The original data resolution was 8.9', these were then smoothed with a Gaussian window of FWHM = 6' producing a map with a final resolution of 10.75'. Additionally, a much lower system temperature was achieved with respect to previous surveys in the same band. For example, S-PASS achieves a factor of two improvement over the Parkes 2.4\,GHz polarisation survey by \citet{Duncan1997}, and they observed only a belt across the Galactic Plane ($|b|<5^\circ$) not covering the area subject of this work. We use the S-PASS Stokes $Q$ and $U$ maps for computation of the RM associated with \bub.

\begin{table*}
\caption{Observational parameters of S-PASS \protect\citep[e.g.][]{Carretti2013} and WMAP K-band \protect\citep{Bennett2013}. * -- Map noise is per observation.} 
	\centering 
	\begin{tabular}{l c c c}
		\toprule 
        Property & Symbol & S-PASS & WMAP\\
		\midrule 
		Reference frequency & $\nu$ & 2307\,MHz & 22.69\,GHz\\ 
		Bandwidth & $\delta\nu$ & 184\,MHz  &  4\,GHz\\ 
        Telescope beamwidth & $\text{FWHM}_\text{tel}$ & 8.9\arcmin  & $0.88\deg$\\
		Map beamwidth & $\text{FWHM}_\text{map}$ & 10.75'  & --\\ 
		Map RMS noise (Stokes $Q$/$U$) & $\sigma$ & $\lesssim$1\,mJy\,beam$^{-1}$   & 1.435\,mK* ($Q$/$U$) -- 6.00\,mK* ($I$)\\ 
		Gain (Jy/K) at $\nu$ & $A$  & $1\,\text{mJy}=0.58\,\text{mK}$  & -- \\ 
        System temperature & $T_\text{sys}$ & $\approx20$\,K  &  29\,K\\
		\bottomrule 
	\end{tabular}
	
	\label{tab:spassparams} 
\end{table*}

\begin{figure*}
	\centering
	\begin{subfigure}{0.49\textwidth}
		\includegraphics[width=\columnwidth]{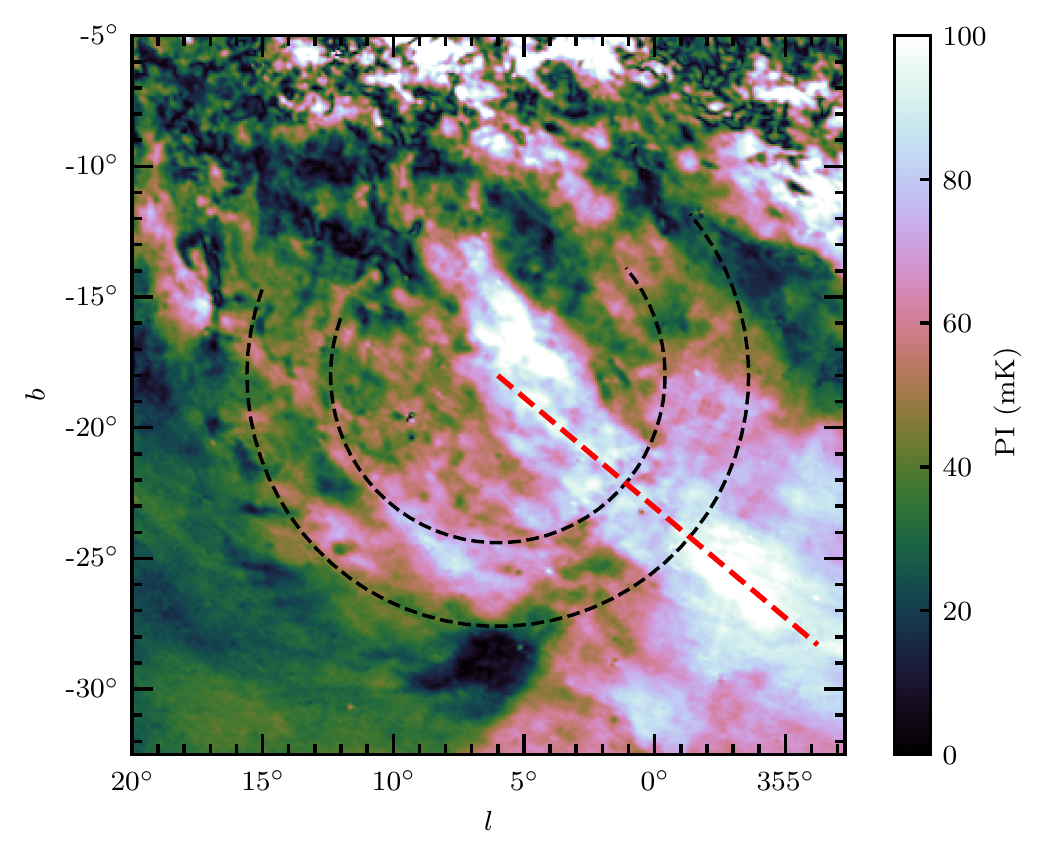}
		\caption{}
		\label{fig:SPASSPIGASS}
	\end{subfigure}
	\begin{subfigure}{0.49\textwidth}
		\includegraphics[width=\columnwidth]{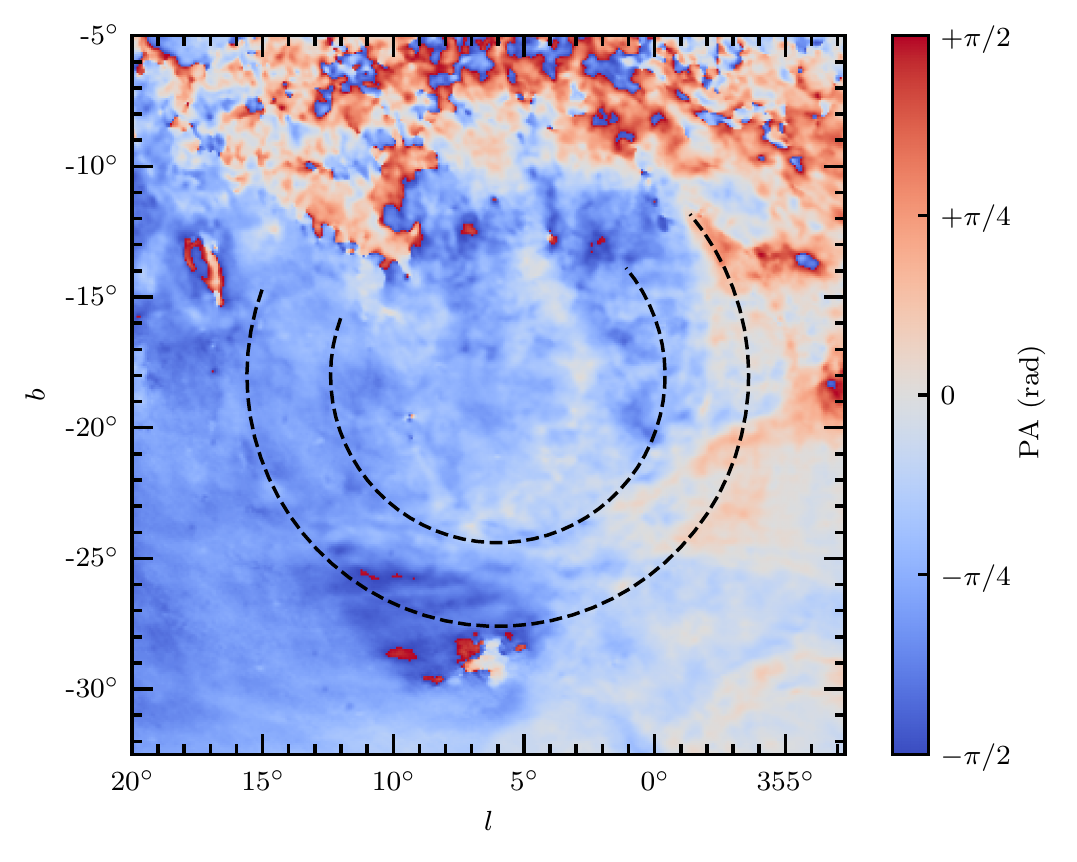}
		\caption{}
		\label{fig:SPASSPAGASS}
	\end{subfigure}
	\begin{subfigure}{0.49\textwidth}
		\includegraphics[width=\columnwidth]{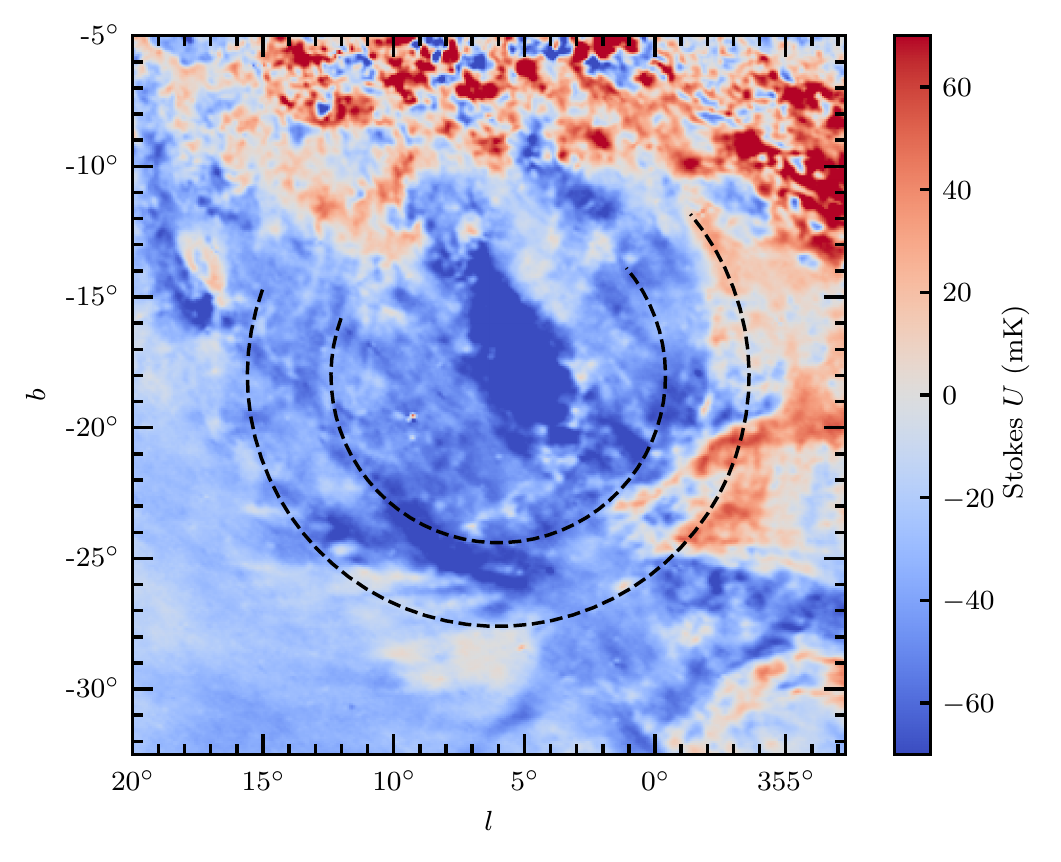}
		\caption{}
		\label{fig:SPASSU}
	\end{subfigure}
	\begin{subfigure}{0.49\textwidth}
		\includegraphics[width=\columnwidth]{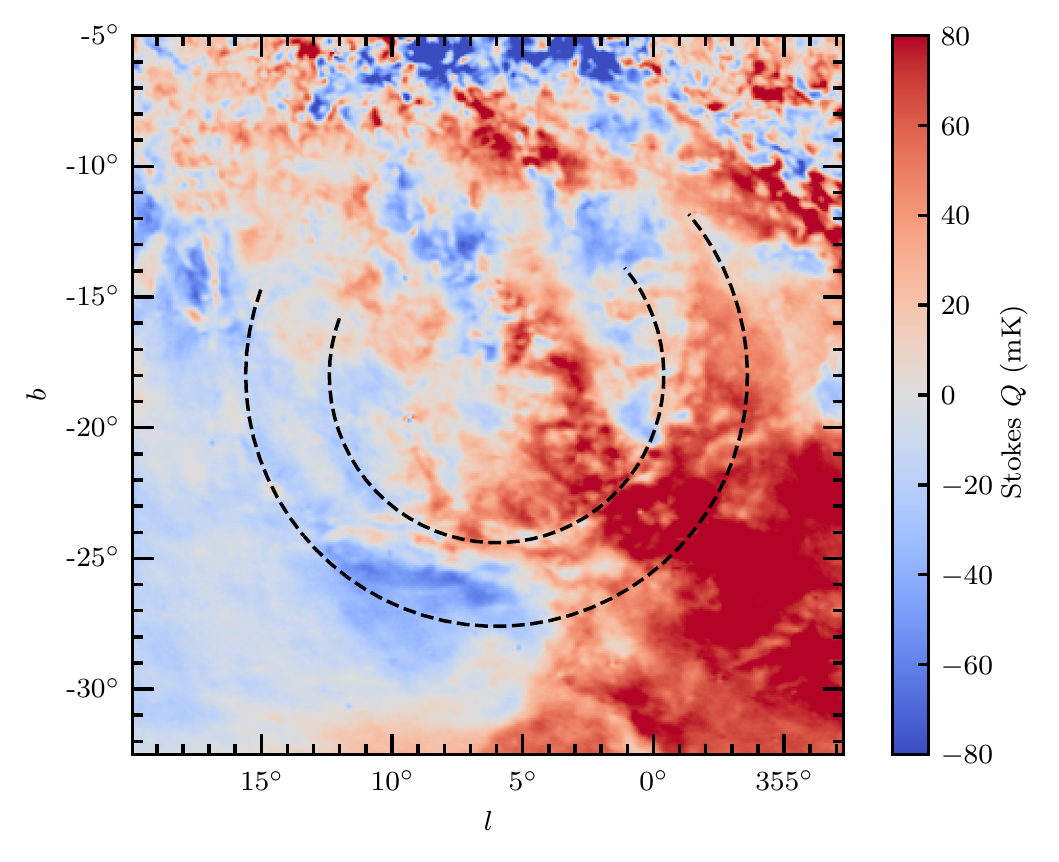}
		\caption{}
		\label{fig:SPASSQ}
	\end{subfigure}
	\caption{Maps of S-PASS polarisation in the region of \bub. \subref{fig:SPASSPIGASS} S-PASS - polarised intensity. The red dashed line corresponds to the profile given in Figure~\ref{fig:profiles}. This profile runs along the ridge of the polarised emission of the Fermi bubble from the centre of the shell region. \subref{fig:SPASSPAGASS} S-PASS - polarisation angle in radians. \subref{fig:SPASSU} S-PASS - Stokes $U$. \subref{fig:SPASSQ} S-PASS - Stokes $Q$. In each panel, the black dashed lines give the approximate inner and outer bounds of the lower part of the shell. }
	\label{fig:images}
\end{figure*}

\subsubsection{Wilkinson Microwave Anisotropy Probe}
The Wilkinson Microwave Anisotropy Probe (WMAP) survey released its final, 9-year data in 2013 \citep{Bennett2013}. The observational properties of WMAP are also summarised in Table~\ref{tab:spassparams}. The project was focused on measuring cosmological parameters and the cosmic microwave background (CMB), however, the foreground data provides invaluable Galactic information. In particular, WMAP K-band, centred on 23\,GHz, gives a calibrated, whole-sky map of polarised synchrotron emission at high frequency. Discussion of the structure present in the WMAP K-band images was also conducted by \citet{Carretti2013}, with a particular focus on comparison with the structure present in the S-PASS data. Specifically, the Northern and Southern Fermi bubbles feature prominently in the polarised emission of WMAP K-band. We make use of the high-frequency polarisation information provided by WMAP and compare these results with S-PASS data in the region of \bub.

\begin{figure}
	\centering
	\begin{subfigure}{0.48\textwidth}
		\includegraphics[width=\columnwidth]{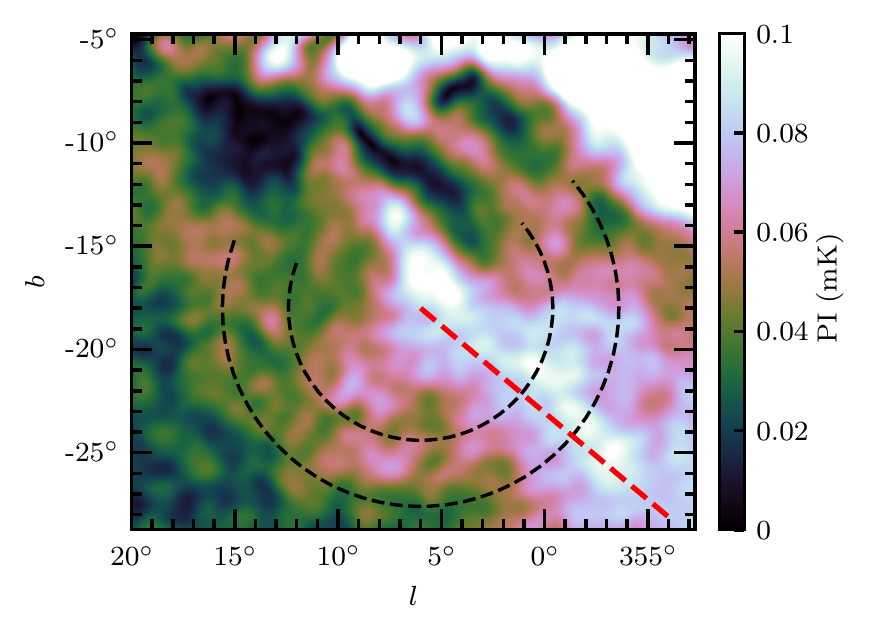}
		\caption{}
		\label{fig:wmappi}
	\end{subfigure}
	\begin{subfigure}{0.48\textwidth}
		\includegraphics[width=\columnwidth]{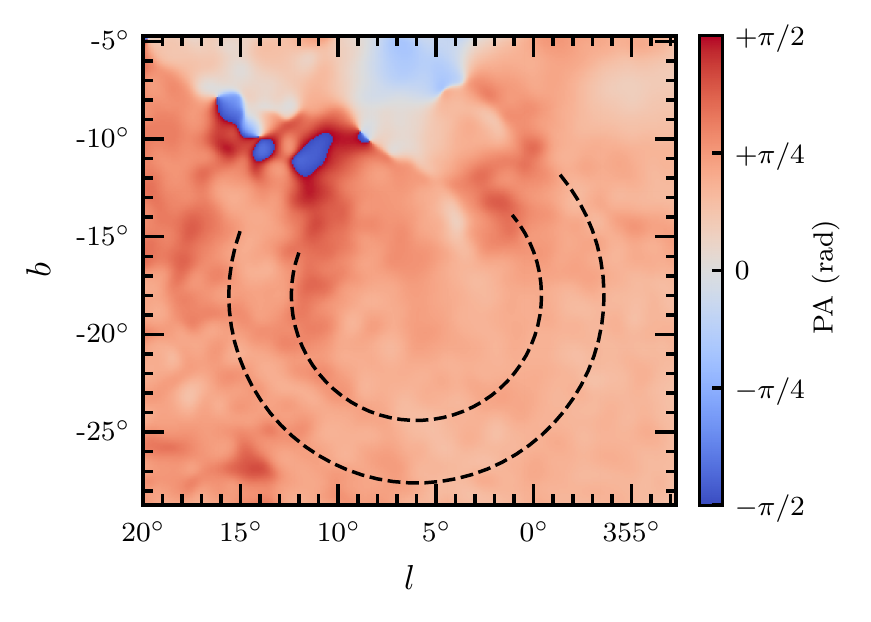}
		\caption{}
		\label{fig:wmappa}
	\end{subfigure}
	\begin{subfigure}{0.48\textwidth}
		\includegraphics[width=\columnwidth]{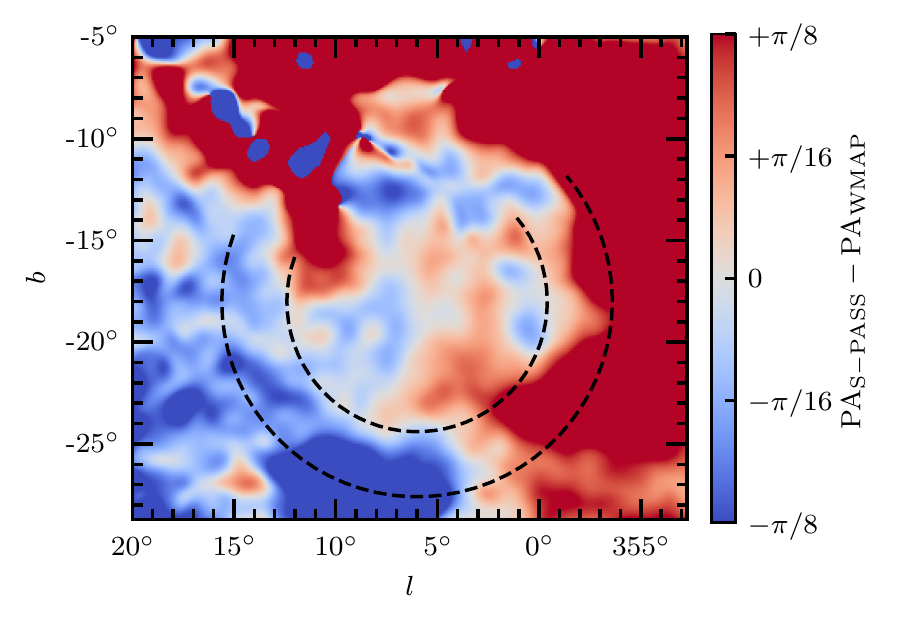}
		\caption{}
		\label{fig:padiff}
	\end{subfigure}
	\caption{ Panels~\subref{fig:wmappi} and \subref{fig:wmappa}: Maps of polarisation in the region of \bub\ at 23\,GHz from WMAP K-band observations. Here we have applied a Gaussian smoothing to the data with a $\text{FWHM}$ of $1\deg$. We applied this filter to increase the signal-to-noise ratio in the WMAP data, which originally had a very high noise level. \subref{fig:wmappi} WMAP - polarised intensity. Again, the red dashed line corresponds to the profile given in Figure~\ref{fig:profiles}. \subref{fig:wmappa} WMAP - polarisation angle in radians. \subref{fig:padiff} Polarisation angle difference between S-PASS and WMAP in radians. Here the S-PASS data were smoothed to common spatial resolution of the smoothed WMAP data. As in Figure~\ref{fig:images}, the dashed lines give the approximate bounds of \bub.}
	\label{fig:wmap}
\end{figure}

\subsection{Radio Continuum}
\subsubsection{Continuum HI Parkes All-Sky Survey}
The Continuum HI Parkes All-Sky Survey (CHIPASS) is a map of the radio continuum at 1.4\,GHz across the whole sky below declination of $\delta=25\deg$~\citep{Calabretta2014}. CHIPASS is a combination and reprocessing of the HI Parkes All-Sky Survey (HIPASS) and the HI Zone of Avoidance (HIZOA) survey, the result of which is a highly sensitive (sensitivity $=40\,$mK), all-sky, total intensity survey at a resolution of 14.4', with very well treated artefacts. We use these data to derive information on the synchrotron spectrum in the region of \bub.

\begin{figure*}
	\centering
	\includegraphics[width=2\columnwidth]{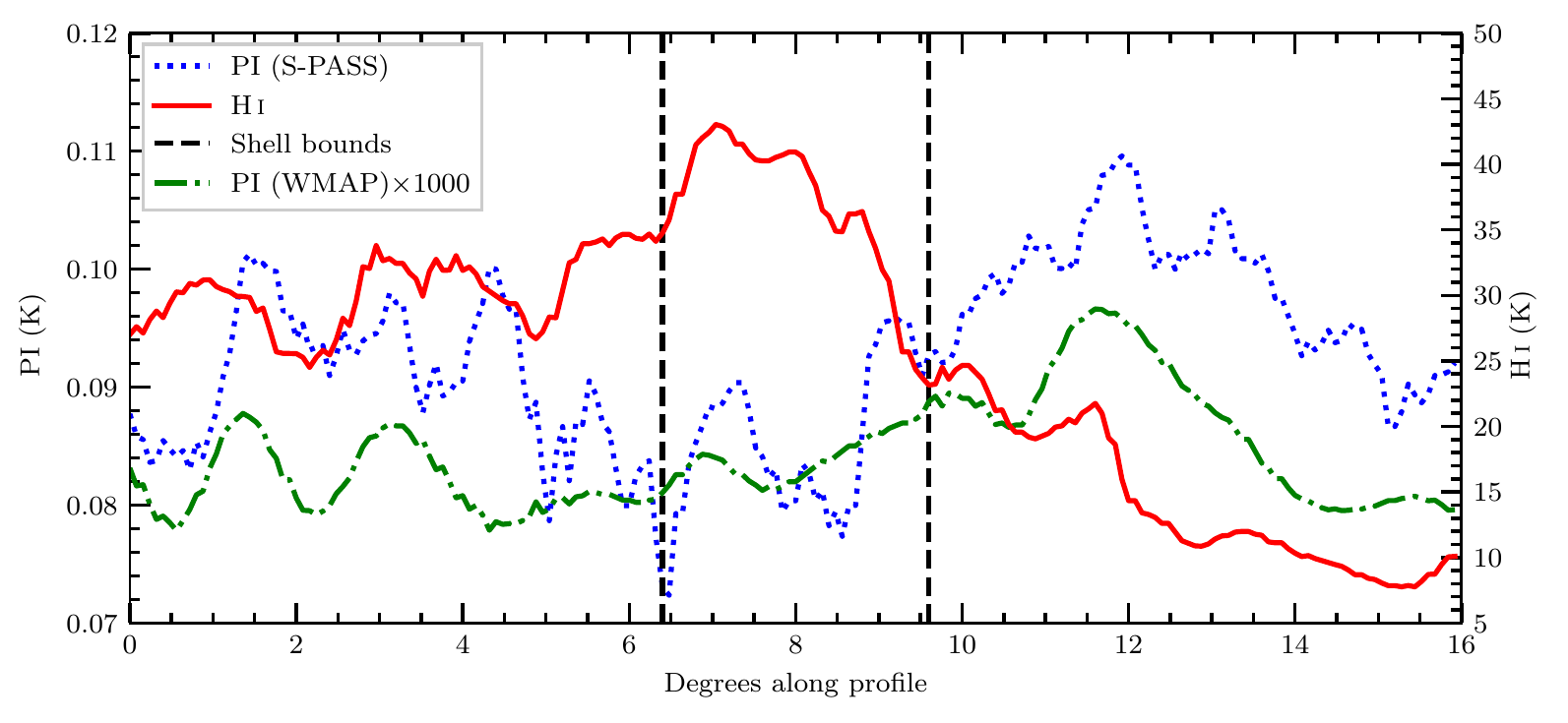}
	\caption{Brightness temperature profiles of polarised intensity (PI) at 2.3\,GHz (blue, dotted), 23\,GHz (green, dash-dotted), and GASS \hi\ at $v_{LSR}=6.6$\,\kms (red, solid). The region from which the profiles are taken is shown in Figures~\ref{fig:GASS} and \ref{fig:images} as the red dashed line. The profile runs from the centre of the shell region to well past the outer boundary, along the bright polarised emission from the Fermi bubble. The black dashed lines correspond to the inner and outer shell boundaries and also shown in Figures~\ref{fig:GASS} and \ref{fig:images}.}
	\label{fig:profiles}
\end{figure*}


\section{Results and Analysis}\label{sec:analysis}
\subsection{Polarisation Morphology}\label{sec:morphology}
The most prominent feature in the diffuse polarisation in the region of \bub\ is the Southern lobe of the Fermi bubbles~\citep{Ackermann2014,Su2010,Carretti2013}, as seen in Figure~\ref{fig:images}. There are number of structures, however, that show significant morphological correlation with the supershell \bub. We claim that these features are perturbations  caused by the MIM of \bub\ to the polarised emission from behind it. This supershell has a distance estimate of $\sim1.5\pm0.5$\,kpc \citep{Moss2012}, and therefore is situated in the foreground relative to the Fermi bubbles (distance of the front surface from Sun $>2.5$\,kpc \citep{Carretti2013}).

The spatial correlation between S-PASS polarisation and \bub, as it appears in \hi\ (see Figure~\ref{fig:GASS}), is most apparent in Stokes parameters $Q$ and $U$. The Stokes $U$ image (see Figure~\ref{fig:SPASSU}) shows the strongest morphological correlation with \bub, although the structure seen in Stokes $U$ is also similar to the structure seen in PA (see Figure~\ref{fig:SPASSPAGASS}). Along the upper and right-hand outer edge of the shell there is a significant shift in the values of Stokes $U$, from $U\approx+0.05\,$K to $U\approx-0.05\,$K. A circular edge can be seen following the right-hand boundary of the shell, where predominately positive $U$ switches to negative inside the boundary. On the left side of the image, where $U$ appears negative, the magnitude of $U$ increases on the shell. There is a bright, polarised region in the centre of the image, which corresponds to the tip of the Fermi bubble. 

The Stokes $Q$ image at 2.3\,GHz (Figure~\ref{fig:SPASSQ}) is dominated by emission from the Fermi bubble. Similar to Stokes $U$, there is a change of sign in $Q$ along the right-hand, inner boundary of the shell as defined by \hi. This region in Stokes $Q$, however, is not as clearly defined as Stokes $U$. Additionally the change of sign occurs along the inner boundary of the shell in $Q$, rather than the outer boundary. A similar feature can be found along the bottom-left, inner boundary of the shell. Here $Q$ again appears to change sign across the inner boundary of the shell. Along the left inner and outer boundaries of the shell we find a weak change in sign of Stokes $Q$.

Inspecting the total linear polarisation intensity (PI) at 2.3\,GHz (Figure~\ref{fig:SPASSPIGASS}) the brightest polarised feature is the Southern ridge of the Fermi bubble \citep{Carretti2013}. This feature runs from the centre of the image to the bottom-right corner. Where the Fermi bubble appears to intersect the shell, however, the polarised intensity is reduced $10\text{-}15\%$ relative to the rest of the lobe. This is indicative that near the boundary of the shell, polarised emission is being perturbed and depolarised. \citet{Carretti2013} presented the first morphological description of S-PASS, focusing particularly on the polarised emission from the Fermi bubbles. They also noted the perturbation feature and suggested that this feature could be due to a line-of-sight reversal of the magnetic field, associated with the Fermi bubbles. We argue however, that due to the high degree of spatial correlation, this feature is explained by the presence of \bub\ in the foreground. This spatial correlation is exemplified in the profiles shown in Figure~\ref{fig:profiles}. Here we can see the anticorrelation of S-PASS PI with \hi\ brightness temperature, especially across the thickness of the shell from $\sim5\deg$ to $\sim10\deg$ along the profile. This demonstrates that the shell is weakly depolarising the background polarised emission from the Fermi bubble. In addition, as depolarisation is a Faraday rotation effect, we also note a change in the polarisation angle across the same boundary in Figure~\ref{fig:SPASSPAGASS}. Depolarisation towards the Galactic plane is also apparent in the PI image, as noted by \citet{Carretti2013}, as well as in Stokes $Q$ and $U$. Significant depolarisation is visible in the region near \bub\ down to a latitude of $b\approx-5\deg$. This region is associated with large \hii\ regions in the Galactic plane, as seen in \halpha\ emission. Additional depolarisation can be seen at latitudes as low as $b\approx-10\deg$. The modulation from this effect makes structure difficult to interpret in the polarisation images near the Galactic plane. 

Unlike the S-PASS polarisation, we see no correlation with \bub\ in any of the WMAP polarisation maps. As seen in Figure~\ref{fig:wmap} the polarised emission at 23\,GHz from the Southern Fermi bubble is the dominant source of polarised emission in the region of \bub. \citet{Carretti2013} also provide a detailed description of the polarised emission from WMAP in this region, again in regards to emission from the Fermi bubbles. We note that there is no noticeable depolarisation across the thickness of the shell, unlike the S-PASS observations. This is particularly clear in Figure~\ref{fig:profiles} where, unlike the S-PASS profile, there is no appreciable drop in the WMAP PI profile as it intersects the shell. The lack of depolarisation is to be expected, as high frequency observations are far less affected by Faraday rotation. We find that the polarisation angle is relatively uniform across the region of \bub\ (see Figure~\ref{fig:wmappa}), with a median value of $\sim37\deg$ and a standard deviation of $\sim9\deg$.  We also note the significant lack of polarised emission in both WMAP and S-PASS in a few regions above the shell along $b\approx-10\deg$. The signal-to-noise in these regions is therefore very low, particularly in WMAP, and also corresponds to large changes in the polarisation angle which are probably spurious. These regions will likely propagate large errors through this analysis.

The strongest morphological indication appears when the polarisation information from S-PASS and WMAP are combined. Inspecting the map of the polarisation angle difference ($\text{PA}_\text{S-PASS}-\text{PA}_\text{WMAP}$) in Figure~\ref{fig:padiff} reveals the circular structure we expect to be associated with \bub. Again, here the right-hand portion of the shell is most prominent, but the entire region is visible in Figure~\ref{fig:padiff} where the angle difference is of relatively low magnitude. This change in polarisation angle is indicative that the shell is Faraday rotating the background polarised emission. The motivation for using this polarisation angle difference is expanded upon in Section~\ref{sec:model}. In short, as the WMAP data was at high frequency, the polarisation as measured should have encountered very little Faraday rotation, in contrast to the S-PASS polarisation. Thus, the difference in polarisation angle contains information on how much Faraday rotation has occurred along the line of sight. As this difference bears a great deal of morphological similarity to \bub, we therefore assume that, within this region, it is the supershell alone that is causing the observed rotation effect.

\subsection{Total Intensity Morphology}
We also carefully inspect the total intensity data from S-PASS, CHIPASS, and WMAP for any morphological indication of \bub. The presence of such emission, especially in the S-PASS data, would mean that the interpretation that the shell was acting as a pure Faraday screen would be false. Faraday screens, by definition, do not produce any emission of their own. We find, however, no such emission in any of these data. The only structure of note is emission from the Corona Australis molecular cloud, visible in WMAP K-band total intensity. This comet-shaped reflection nebula is a common feature in infra-red observations \citep{Schlegel1998,Neuhauser2008}, and appears to lie along the bottom edge of \bub. However, its overall morphology is very distinct from \bub, and sits far closer to us at a distance of $\sim130\,$pc \citep{Neuhauser2008}.


\subsection{Faraday Screen Model}\label{sec:model}
\subsubsection{General concept of the Faraday screen model}
\begin{figure}
	\centering
	\def\svgwidth{\columnwidth}
	{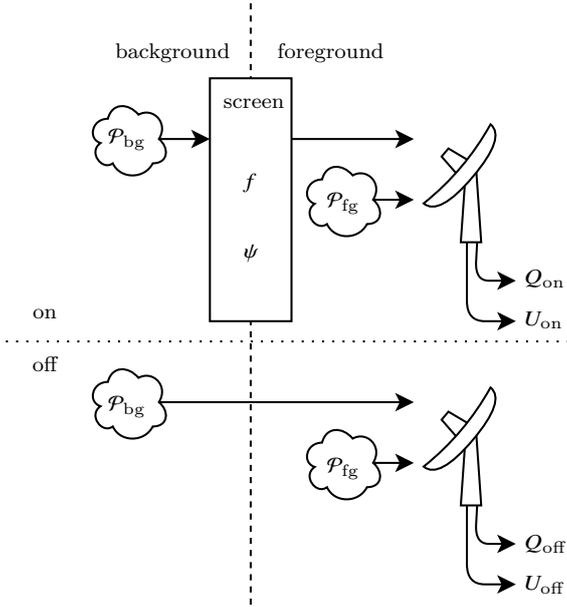}
	\caption{Cartoon of the Faraday screen model. The `foreground'/`background' divide refers to the line-of-sight region where the polarised emission is produced, relative to the screen. The `on'/`off' divide refers to whether the line-of-sight observes through the screen or not. In both the `on' and `off' case the measured polarisation ($\mathcal{P}$) is the superposition of the `background' and `foreground' emission. In the `on' case, however, the background emission is  perturbed by the screen.}
	\label{fig:cartoon}
\end{figure}
As discussed above, \bub\ does present a clear polarisation signature at 2.3\,GHz. This signature however is not directly evident as polarised emission, nor is there a clear correlation in \halpha\ emission \citep{Moss2012}. This is indicative of a relatively low free electron density (compared to an \hii\ region). This type of interaction is dubbed a `Faraday screen' effect \citep{Wolleben2004,Sun2007a}. Faraday screens are simple Faraday rotating regions that affect the synchrotron emission that is produced from behind them (along the line of sight). Faraday screens do not produce polarised emission themselves, instead they rotate the background polarisation angle, and/or reduce the polarised intensity through depolarisation effects. \citet{Sun2007a} provide a model for determining the Faraday rotation that occurs due to the presence of a Faraday screen. This Faraday screen model is visualised in Figure~\ref{fig:cartoon}. In this scenario, an observer can measure polarisation ($\mathcal{P}$) either `on' or `off' the screen. The terms `on' and `off' refer to whether the polarised emission has been affected by the Faraday screen or not, respectively. In either case the model assumes that the observer will measure the superposition of the polarised emission from the `background' and `foreground', relative to the screen. When observing `on' the screen the background polarised emission will be perturbed. Specifically, the polarised intensity will be multiplied by the depolarisation factor `$f$' (where $f<1$), and the polarisation angle will be rotated by an amount `$\psi$'. As such, the Stokes parameters `on' the screen are given by \citep[from][]{Sun2007a}:

\begin{equation}
\begin{aligned}
U_{\text{on}} &= \text{PI}_{\text{fg}}\sin{(2\psi_0)}+f\text{PI}_{\text{bg}}\sin{[2\left( \psi_0 + \psi\right)] }\\
Q_{\text{on}} &= \text{PI}_{\text{fg}}\cos{(2\psi_0)}+f\text{PI}_{\text{bg}}\cos{[2\left( \psi_0 + \psi\right)] }
\end{aligned}
\label{eqn:onoff}
\end{equation}
These equations assume that the intrinsic polarisation angles in the background and foreground are related by $\text{PA}_\text{bg}=\text{PA}_\text{fg}=\psi_0$. This assumption is in contrast to \citet{Sun2007a} who assumed that $\psi_0\approx0\deg$. We are unable to make this same assumption, as it implies that $\text{PA}_\text{off}\approx0\deg$. If we take the WMAP polarisation angle information as an estimate of $\text{PA}_\text{off}$, it is clear from Figure~\ref{fig:wmappa} that this is not the case. Our assumption is reasonable in the case of either: 1) a random field, or 2) a dominant coherent field in a single direction. From this, we derive the same model of a Faraday screen as \citet{Sun2007a}, using observations of polarised intensity (PI) and polarisation angle (PA), `on' and `off' the screen:

\begin{equation}
\begin{aligned}
&\frac{\text{PI}_\text{on}}{\text{PI}_\text{off}}=\sqrt{f^2(1-c)^2+c^2+2fc(1-c)\cos{2\psi}}\\
&\text{PA}_\text{on}-\text{PA}_\text{off}=\frac{1}{2}\arctan{\left(\frac{f(1-c)\sin{2\psi}}{c+f(1-c)\cos{2\psi}}\right)}
\end{aligned}
\label{eqn:screen}
\end{equation}
This model describes the change in polarised intensity and polarisation angle between observations `on' and `off' the screen. Here the four observables are $\text{PI}_\text{on}$, $\text{PI}_\text{off}$, $\text{PA}_\text{on}$, and $\text{PA}_\text{off}$. The model parameters are $f$, $c$, and $\psi$. The parameters $f$ and $\psi$ are as described above, where $f$ is the factor of depolarisation that has occurred at the observed frequency, with $f\in[0,1]$, while $\psi$ is the amount of Faraday rotation through the screen, with $\psi\in[-\pi/2,+\pi/2]$. The parameter $c$ is the fraction of foreground polarised intensity given by: 
\begin{equation}
	c=\frac{{\text{PI}}_\text{fg}}{\left(\text{PI}_\text{fg}+\text{PI}_\text{bg}\right)}
	\label{eqn:c}
\end{equation}

The range of this parameter is therefore $c\in[0,1]$. The Faraday thickness ($\phi$) of the screen can be estimated from this model by: 
\begin{equation}
\phi=\frac{\psi}{\lambda^2}
\label{eqn:psi}
\end{equation}

Here `Faraday thickness' refers to the Faraday depth of the shell alone. This equation applies in the case of a rotating-only Faraday screen, as the RM of the screen is equal to the Faraday depth. This model has an advantage over the model presented in \citet{Wolleben2004}, as it only requires observations at a single frequency.

In previous uses of this model the `on' and `off' components were taken from a common set of radio polarisation observations at the same frequency, but were spatially offset to sample `on' and `off' the proposed location of the Faraday screen \citep{Sun2007a,Sun2011,Gao2010,Gao2015}. There are two primary assumptions made by this technique: (1) that the polarised emission located beside (`off') the screen region accurately represents the superposition of synchrotron emission from the foreground and background; and, (2) that all Faraday rotation along the line of sight occurs at the Faraday screen. In our analysis, the Faraday screen region of \bub\ appears in the direction of the Galactic plane. As the line of sight approaches the plane, the amount of Faraday rotating structure increases; further, the angular extent of \bub\ is enormous. As such, assumption (1) is not easily satisfied here.

\subsubsection{Using WMAP to obtain `off' polarisation}
In contrast to previous uses of the \citet{Sun2007a} Faraday screen model, we use S-band-scaled, high frequency (23\,GHz), polarisation data as the `off' measurements to the S-PASS `on' measurements. By utilising polarisation data from a high frequency survey we bypass the need for assumption (1) to be satisfied. That is, rather than using a spatially-offset pointing at the same frequency for the `off' measurement, we use pointings towards the screen, but at a different frequency that is negligibly affected by the screen, thus serving as the `off' measurement. Additionally, this allows for the computation of a Faraday depth for each line of sight through the Faraday screen. 

With this consideration, we obtain high frequency polarisation data from the WMAP survey. As Faraday rotation increases as $\lambda^2$, only strong Faraday depth sources ($\phi>100\,$\radms) can produce non-negligible Faraday rotation at 23\,GHz. Since \bub\ is an object in the diffuse ISM, likely occupying some combination of the warm neutral and warm ionised medium \citep{Heiles2012}, we do not expect significant Faraday thickness to be produced by this object.

A limitation of a Faraday screen model is that this technique is only sensitive to Faraday rotation values of $-90\deg<\psi<+90\deg$, due to assumption (2). Any rotations greater than this through the screen would create `$n\pi$' ambiguities in the measured polarisation angle. As such, we are only able to obtain Faraday depths between $\phi=\pm{({\pi}/{2\lambda^2})}\approx\pm92\,$\radms. We can exclude the possibility of `$n\pi$' ambiguities by inspecting the morphology of the polarisation angle image from S-PASS (see Figure~\ref{fig:SPASSPAGASS}). Regions that strongly Faraday rotate appear to have `onion skin' structure in polarisation angle. That is, across these regions the polarisation angle changes sign multiple times towards the centre of the object. This corresponds to multiple revolutions of the polarisation vector as the physical and Faraday depth of the object increases towards the centre. The \bub\ region does not exhibit this morphology, rather it has an approximately uniform polarisation angle structure within the boundaries of the shell. This is therefore consistent with the assumption that the amount of rotation through the shell is $<|90\deg|$. 

WMAP K-band does suffer low signal-to-noise ratio as it measures synchrotron radiation at the far end of the spectrum. The low signal-to-noise of the WMAP data must be addressed in order to be used in analysis of the Galactic synchrotron emission. We do this by applying Gaussian smoothing to these images. We convolve a Gaussian smoothing function with a $\text{FWHM}=1\deg$ with both the Stokes $Q$ and $U$ maps. 

\subsubsection{Frequency scaling of WMAP to S-Band}
In order to estimate the synchrotron background at S-Band, we need to scale WMAP data to 2.3GHz. To do this we require the synchrotron spectral index ($\beta$) between these frequencies. Between any two frequencies ($\nu$) with total intensity ($T$), the spectral index is given by:
\begin{equation}
	\beta = \frac{\log{\left(T_1/T_2\right)} }{\log{\left( \nu_1/\nu_2\right) }}
	\label{eqn:beta}
\end{equation}
As the zero--offset calibration total intensity map of S-PASS is not yet finalised, we instead obtain the synchrotron spectral index between 1.4\,GHz and 23\,GHz. This assumes a spectrally constant $\beta$ between these frequencies. We then use this value of $\beta$ to scale the WMAP K-band data to S-band. To find the spectral index we use the total intensity data at 1.4\,GHz from the CHIPASS survey \citep{Calabretta2014}. The synchrotron spectral index can usually be obtained from the slope ($m$) of the T-T plot of the total intensity (Stokes $I$) between two frequencies using \citep{Turtle1962}:
\begin{equation}
	\beta\approx\frac{\log{m}}{\log{\left({\nu_\text{W}}/{\nu_\text{C}}\right)}}
\end{equation}
where `W' stands for WMAP, `C' for CHIPASS. We find, however, that the slope of the T-T plot is not constant in the region of \bub, indicating thermal emission processes (see Section~\ref{sec:tt}).

\begin{figure}
	\centering
	\includegraphics[width=\columnwidth]{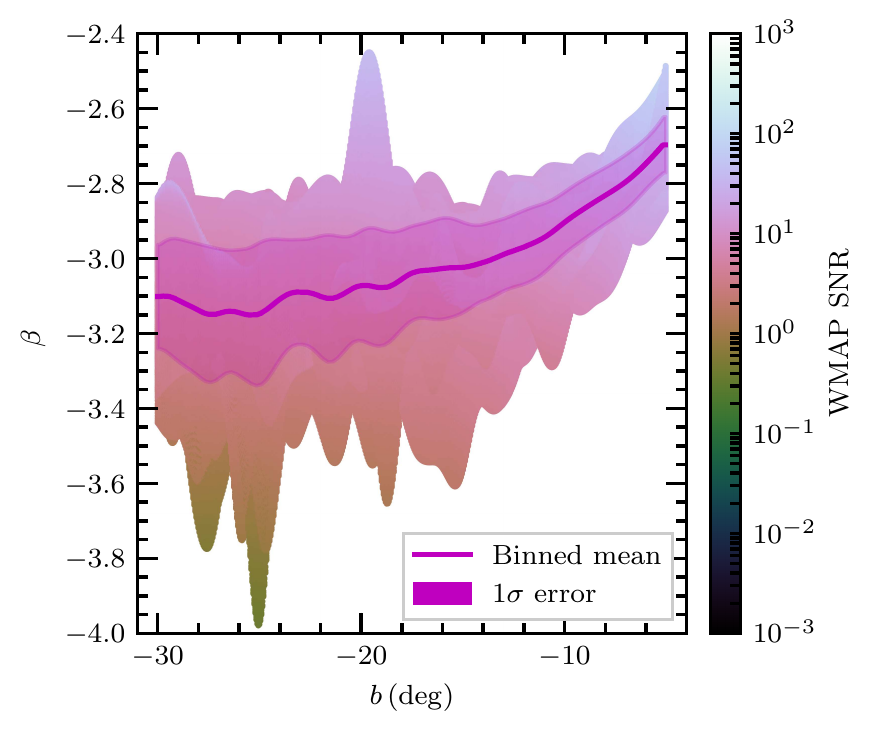}
	\caption{Spectral index for each point in 
     the region of \bub\ as a function of Galactic latitude. The longitude range used is $45\deg<l<315\deg$. The points are coloured by the signal-to-noise ratio (SNR) of WMAP Stokes $I$. Despite the smoothing applied to the data, the effects of point sources and regions of low intensity are still present. These effects cause a large variance in the derived spectral index. However, after we apply a box-car smooth to the data, we recover a consistent spectral index of $\beta\sim-3.1$ in the region of the shell. Additionally, we also find a clear trend of the spectral index flattening towards the Galactic plane.}
	\label{fig:beta}
\end{figure}

Rather than using this result, we compute the spectral index for each point in the region of \bub\ using Equation~\ref{eqn:beta}. We then analyse the indices as a function of latitude, as shown in Figure~\ref{fig:beta}. The spectral index remains relatively constant for $b<-10\deg$; after which $\beta$ starts to increase to values greater than $-3$, as also indicated in the T-T plot. Following this, we bin the obtained indices, and compute the error from this binning (shown as the solid line with the 1$\sigma$ uncertainty band in Figure~\ref{fig:beta}). Based on Figure~\ref{fig:beta}, we use the binned spectral index as a function of Galactic latitude and assume that it is constant in Galactic longitude. We use this to scale WMAP $Q$ and $U$ to S-Band, and carry the associated errors forward. The presence of thermal contamination in the spectral index with $b\gtrsim-10\deg$ is irrelevant for the following analyses, because the region of \bub\ that is depolarised in S-band is also at $b\gtrsim-10\deg$ and is thus excluded. We note that using this technique requires that the CMB emission must be subtracted from CHIPASS, as it is not present in the WMAP K-band data. A T-T plot technique usually side-steps this requirement, but since we are unable to use such a technique in this case we subtract 2.7\,K uniformly from the CHIPASS total intensity data.

\subsubsection{Summary of our Faraday screen observables}
Finally, we obtain the four observables, as required for the Faraday screen model, as follows:
\begin{equation}
	\begin{aligned}
		\text{PI}_\text{on} &= \sqrt{Q_\text{S}^2+U_\text{S}^2}\\
		\text{PI}_\text{off} &= \left( \frac{\nu_\text{S}}{\nu_\text{W}}\right)^\beta\sqrt{\left(Q_\text{W}\right)^2 + \left(U_\text{W}\right) ^2}\\
		\text{PA}_\text{on} &= \frac{1}{2}\arctan{\left( \frac{U_\text{S}}{Q_\text{S}}\right) }\\
		\text{PA}_\text{off} &= \frac{1}{2}\arctan{\left( \frac{U_\text{W}}{Q_\text{W}}\right) }
	\end{aligned}
	\label{eqn:inputs}
\end{equation}
here $Q$ and $U$ refer to the Stokes $Q$ and $U$ maps from S-PASS (`S') and WMAP (`W'), $\beta$ is the spectral index binned as a function of latitude (see Figure~\ref{fig:beta} above), and $\nu$ refers to the frequency of each survey. We note there are some regions that have $\text{PI}_\text{on}>\text{PI}_\text{off}$, which cannot be accommodated by the model. These regions will be excluded from later analysis.

\subsection{Best Fit Procedure}\label{sec:bestfit}
\begin{figure}
	\centering
	\includegraphics[width=\columnwidth]{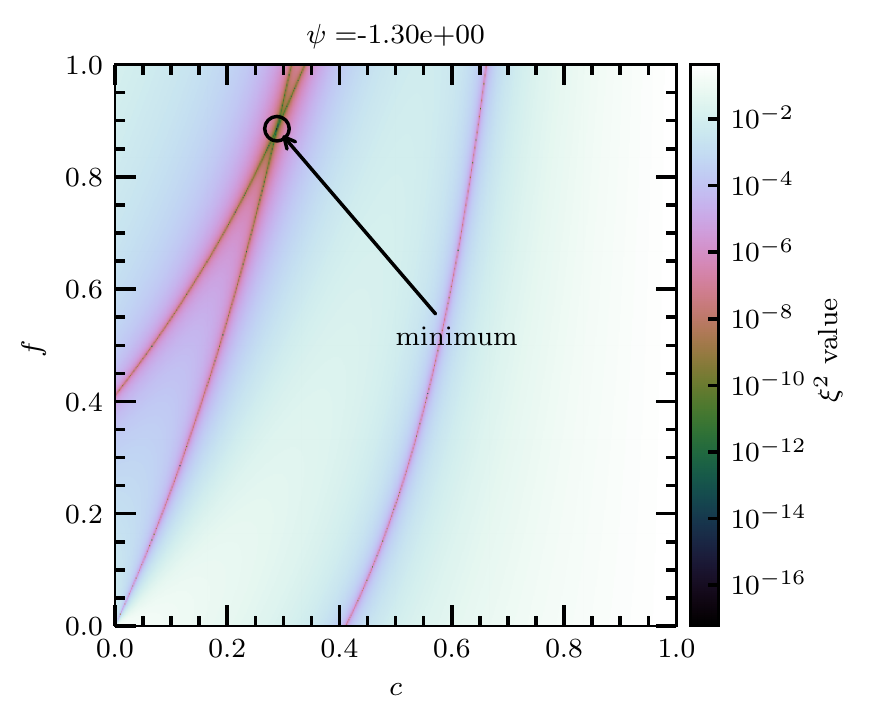}
	\caption{Slice through $\xi^2$ cube for a random line-of-sight ($l,b=[0.0\deg,-24.96\deg]$) at mean input value. The cube shown is a $500^3$ grid, whose $[x,y,z]$ coordinates correspond to the parameters $[f,c,\psi]$. The cube is sliced through the $\psi$ coordinate at the minimum value of $\xi^2$. Note that in our final analysis, we use a $50^3$ grid, which still provides robust results.}
	\label{fig:cube}
\end{figure}

From the Faraday screen model (Equation~\ref{eqn:screen}), we have four observables and three model parameters. We obtain the observables from the data using Equation~\ref{eqn:inputs}. To obtain the model parameters we need to fit the Faraday screen model to the data. In order to determine the `best fit' of the Faraday screen model, we have developed a `brute-force' method that determines which values of the parameters $f$, $c$, and $\psi$ produce the closest fit to the input data. To this end, we implement a grid search technique of the model hyperspace. Specifically, we sample the model for a large number of possible parameter values and find the combination of the parameter values that provide the closest fit to the data. For the purposes of our analysis, each parameter is allowed to take values in the ranges $f\in[0,1]$, $c\in[0,1]$, and $\psi\in[-\frac{\pi}{2},+\frac{\pi}{2}]$. We refer to the number of samples of parameters $f$, $c$, and $\psi$ as $N_f$, $N_c$, and $N_\psi$, respectively. For simplicity we use the same number of samples for each parameter; namely $N_f=N_c=N_\psi=N_\text{samp}$. This is possible because the final results are insensitive to the number of samples per parameter, provided a large enough value is used. We then evaluate the Faraday screen model (Equation~\ref{eqn:screen}) for all of these parameter values. This results in two hypersurface cubes of size $N_{\text{samp}}^3$; one containing all the possible model (mod) PI ratio ($\text{PIR}=\text{PI}_\text{on}/\text{PI}_\text{off}$) values and the other the model PA difference ($\text{PAD}=\text{PA}_\text{on}-\text{PA}_\text{off}$) values. For a given set of observed (obs) PIR and PAD we find the following `$\xi^2$' quantity:
\begin{equation}
	\xi^2 = \left[ \text{PIR}_\text{obs}-\text{PIR}_\text{mod}\right]^2\times\left[ \text{PAD}_\text{obs}-\text{PAD}_\text{mod}\right]^2
\end{equation}
This quantity is constructed similarly to $\chi^2$, but with a few key differences. Each component in $\xi^2$ is not weighted by the variance from an underlying distribution. This is because the observed values are themselves sampled from a probability distribution function (PDF) which we produce. In principle we could compute the variance from this distribution and then find and minimise the $\chi^2$. In doing so, however, we would lose the ability to propagate a PDF through the modelling process. As the components of $\xi^2$ are not weighted by the variance, each factor still has the dimensions of the observable. That is, we have the dimensionless PIR quantity and the PAD quantity in radians. Thus, unlike $\chi^2$, we take the product of the two factors. Like $\chi^2$ the smallest value within our $\xi^2$ cube corresponds to the values of $f$, $c$, and $\psi$ which provide the best fit to the data. This minimum value should also correspond to $\xi^2=0$, provided a solution exists, within numerical precision. We check for multiple solutions of ($ f $, $ c $, $ \psi $) that could produce $\xi^2=0$, but we only ever find a unique solution for ($ f $, $ c $, $ \psi $) for which $\xi^2=0$. Additionally, we check for the existence of an exact solution for each input pixel, and exclude pixels for which no solution exists from further analysis.  We provide an example of where we find a minimum in the $\xi^2$ cube for a randomly sampled line-of-sight input in Figure~\ref{fig:cube}. This example is typical of how the hypersurface appears; with the location of the minimum in the cube occurring at the intersection of two `troughs' of local minima, and being numerically very close to 0.

Our method does not rely on choosing reasonable initial conditions, nor does it require iteration for convergence. Instead, once the grid is built, the best fit is immediately obtained in the ($ f $, $ c $, $ \psi $) parameter grid as the location of the global minimum within the $\xi^2$ cube.

To obtain the error in the fit of this model we apply a Monte-Carlo approach. First, we obtain the PDF of each input parameter, assuming that the starting input data (S-PASS Stokes $Q$ and $U$, WMAP Stokes $Q$ and $U$, and $\beta$) have Gaussian distributed errors. For each pixel in each of the input data maps we produce a Gaussian PDF. To compute the PDFs of the observables, as required for the Faraday screen model, we apply the arithmetic described in Equation~\ref{eqn:inputs} to each value in each PDF. These PDFs become the priors, which we input into the best-fit algorithm; where we find a best-fit for the entire prior PDF for every pixel in the region around \bub. We sample the PDFs as a histogram with $N_\text{PDF}$ bins. 

By propagating the entire prior PDF through the best-fit process we are able to obtain the full posterior PDF of the model parameters $f$, $c$, and $\psi$. Thus, we can gain an estimation of the uncertainty in each parameter. Examples of these posterior PDFs are given in Figure~\ref{fig:post}. We obtain the number of samples, $N_{\text{PDF}}$ and $N_\text{samp}$, from a convergence test on the first moment values of random pixels in the region of \bub. We tested values of $N_{\text{PDF}}$ between 100 and $10^6$, and $N_\text{samp}$ between 1 and 200. We found $N_{\text{PDF}}=3000$ and $N_\text{samp}=50$ to be the minimum values that still provide robust posterior results. For the entirety of our analysis we use these two sample values across the entire region.

\begin{figure}
	\centering
	\begin{subfigure}{0.49\textwidth}
		\includegraphics[width=\columnwidth]{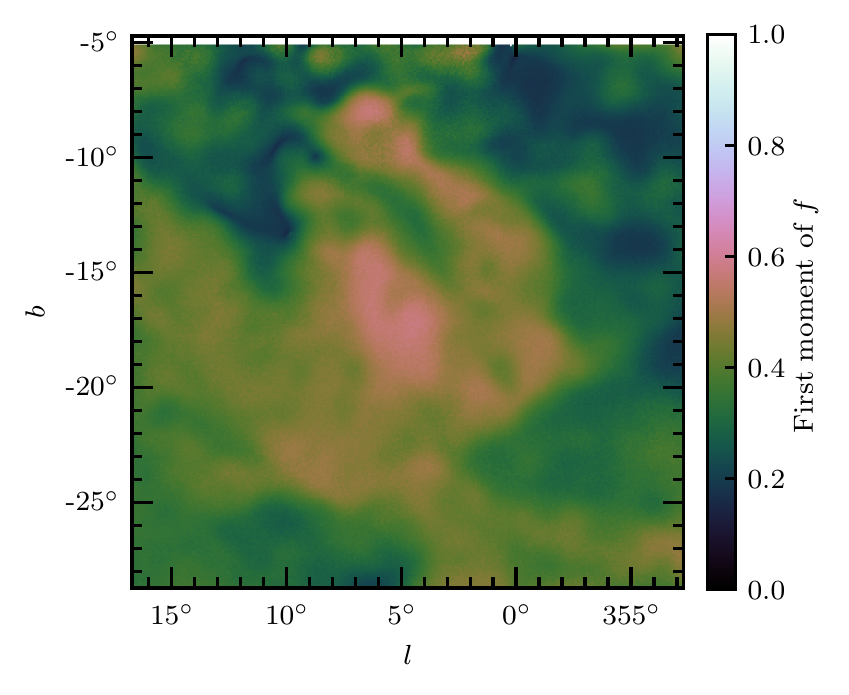}
		\caption*{}
		\label{fig:fmap}
	\end{subfigure}
	\begin{subfigure}{0.49\textwidth}
		\includegraphics[width=\columnwidth]{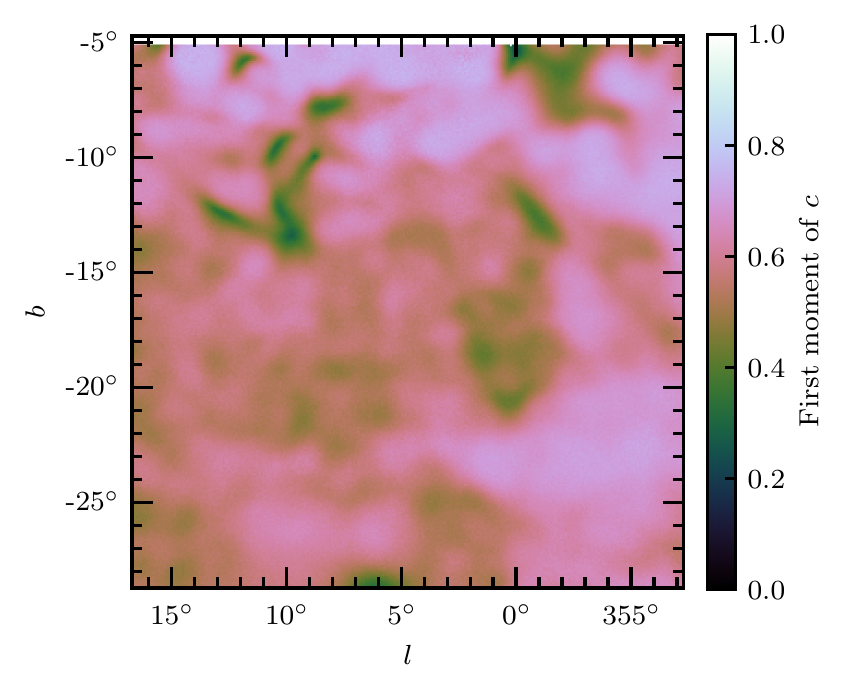}
		\caption*{}
		\label{fig:cmap}
	\end{subfigure}
	\caption{First moment map of model parameters in the region of \bub: the depolarisation factor, $f$ (upper panel), and the fraction of foreground polarisation, $c$ (lower panel).}
	\label{fig:modelmaps}
\end{figure}

\begin{figure}
	\centering
	\begin{subfigure}{0.49\textwidth}
		\includegraphics[width=\columnwidth]{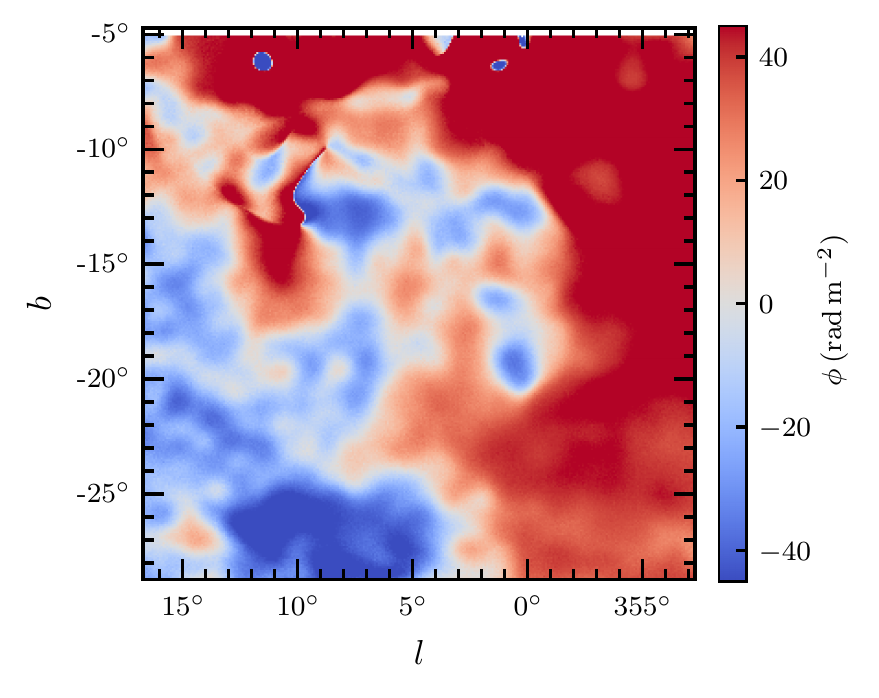}
		\caption*{}
		\label{fig:depthmap}
	\end{subfigure}
	\begin{subfigure}{0.49\textwidth}
		\includegraphics[width=\columnwidth]{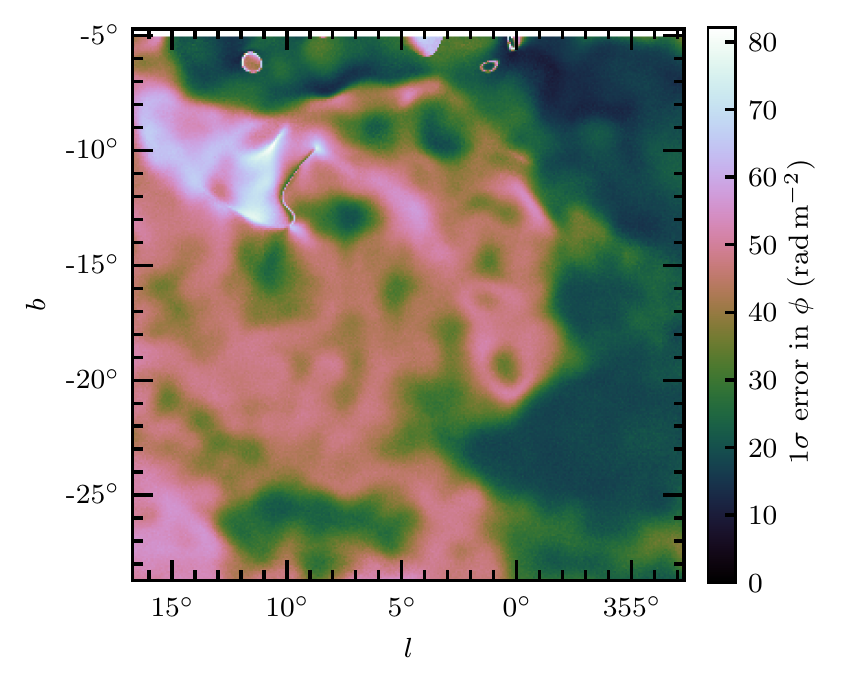}
		\caption*{}
		\label{fig:error}
	\end{subfigure}
	\caption{First moment map of Faraday depth in the region of \bub\ (upper panel) and $1\sigma$ error associated with this Faraday depth from the model fit (lower panel).}
	\label{fig:faradaymap}
\end{figure}

This process is repeated for each pixel available in the S-PASS map in the vicinity of \bub; the result of which is a PDF for each model parameter ($f$, $c$, and $\psi$) for each pixel on the map. Additionally, we also produce the best-fit values from a $\chi^2$ minimisation for comparison with the $\xi^2$ results. We find that the model values from this $\chi^2$ minimisation are consistent with centroid values from the distributions of the model parameters. We present here a map of the first moment of the model parameters $f$ and $c$ (Figure~\ref{fig:modelmaps}). We also present a map of the first moment of Faraday thickness, as given by Equation~\ref{eqn:psi}, and its $1\sigma$ error map (Figure~\ref{fig:faradaymap}). 

\begin{figure}
	\centering
	\includegraphics[width=\columnwidth]{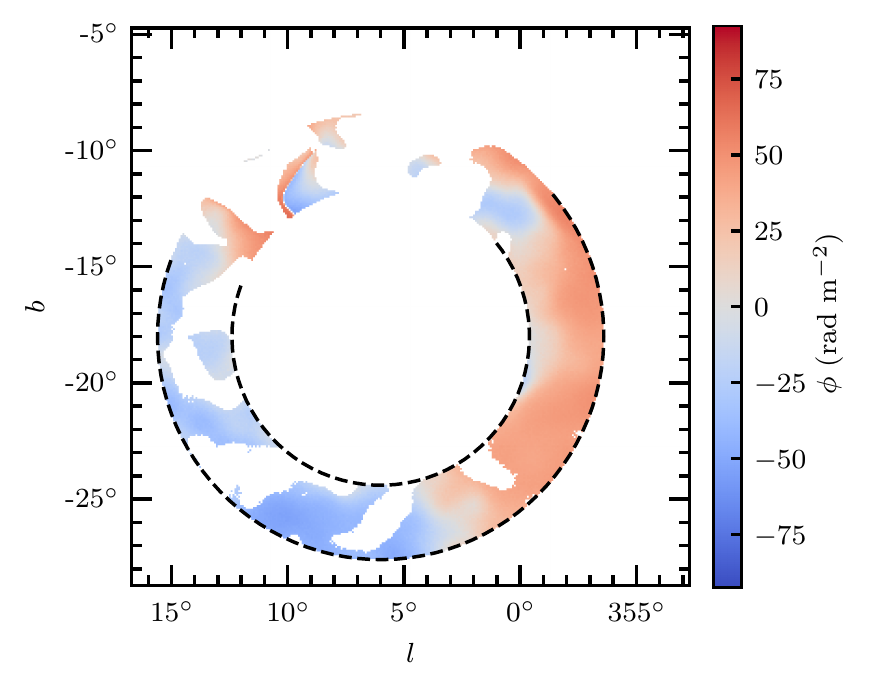}
	\caption{First moment map of Faraday thickness of \bub. A mask was applied to remove regions where the Faraday screen model does not apply.}
	\label{fig:faradaymask}
\end{figure}

We exclude regions in the output data where the Faraday screen model does not apply i.e. where $\text{PIR}_\text{obs}>1$, and regions outside the outer boundary of \bub\ or inside of the inner boundary, as defined by its \hi\ emission. The resulting map is given in Figure~\ref{fig:faradaymask}. We also apply the same mask to the maps of $f$ and $c$. Within this masked region both $f$ and $c$ have a roughly uniform value, with spatial standard deviations of 0.08 and 0.06, respectively. We therefore compute the mean values of $f$ and $c$ and their associated errors. From this we find the average depolarisation factor $f_\text{av}=0.4\pm0.3$ and fraction of foreground polarisation $c_\text{av}=0.6\pm0.3$. The errors given here are derived from the uncertainties in the mean, and not from the spatial variation. These values imply that on average roughly half of the background emission is depolarised by \bub, and that the background and foreground polarised emission are approximately on parity, with the foreground being slightly dominant.

\subsection{Error Propagation of Best-Fit Parameters}
\begin{figure}
	\centering
	\begin{subfigure}{0.49\textwidth}
		\includegraphics[width=\columnwidth]{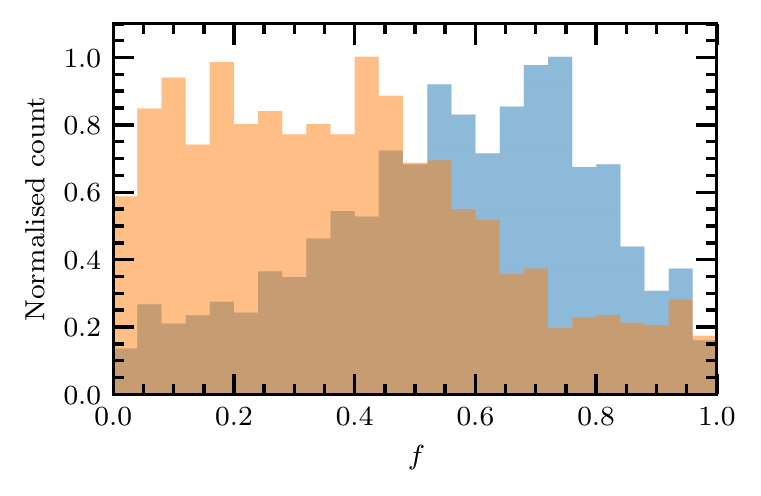}
		\caption*{}
		\label{fig:postf}
	\end{subfigure}
	\begin{subfigure}{0.49\textwidth}
		\includegraphics[width=\columnwidth]{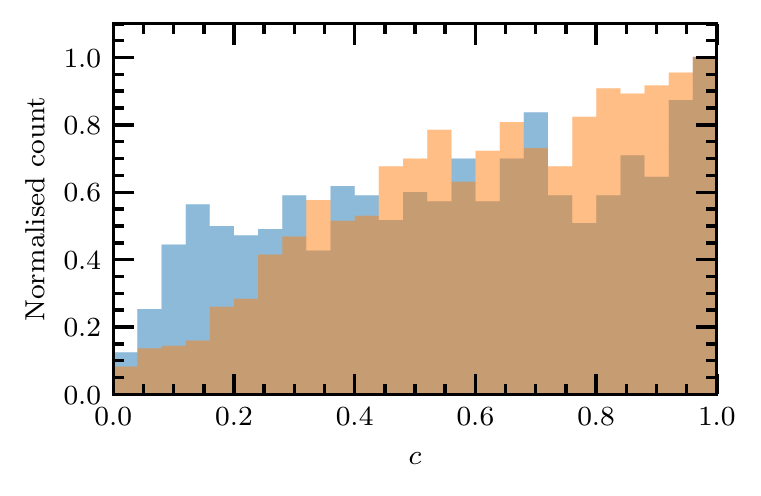}
		\caption*{}
		\label{fig:postc}
	\end{subfigure}
	\begin{subfigure}{0.49\textwidth}
		\includegraphics[width=\columnwidth]{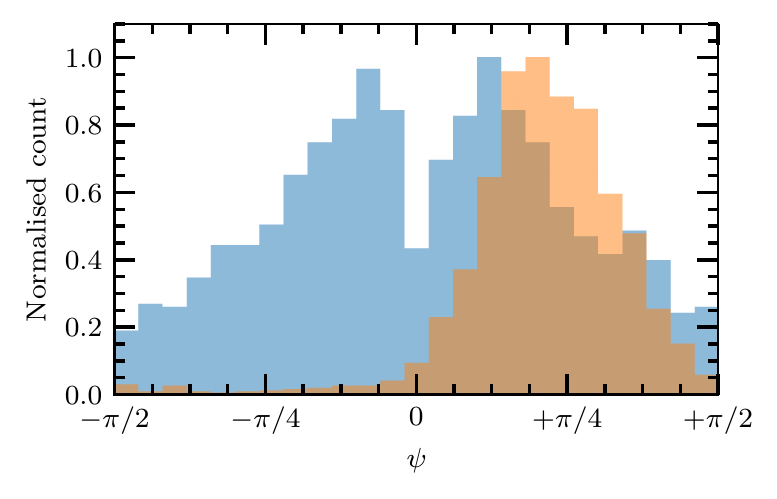}
		\caption*{}
		\label{fig:postp}
	\end{subfigure}
	\caption{Posterior PDFs for two lines-of-sight in the $f$, $c$, and $\psi$ data. Blue: $l,b = [4.96\deg,-18.0\deg]$, Orange: $l,b=[0.0\deg,-24.96\deg]$. Recall that $f$ is the depolarisation factor, $c$ is the fraction of foreground polarisation, and $\psi$ is the amount of Faraday rotation through the screen. } 
	\label{fig:post}
\end{figure}

We use a Monte Carlo error propagation method to carry the errors forward through our best-fit procedure. As such, we are able to constrain regions within the output data where the fit is poor. Figure~\ref{fig:post} shows the PDFs of $f$, $c$, and $\psi$ for two pixels sampled at $l,b = [4.96\deg,-18.0\deg]$ (blue area) and $l,b=[0.0\deg,-24.96\deg]$ (orange area). Recall $f$ is the depolarisation factor, where a value of 0 refers to total depolarisation, and a value of 1 to no depolarisation. The parameter $c$ is the fraction of foreground polarisation, where values of 0 implies that all the emission is from the background, and 1 implies that all the emission originated in the foreground. These two pixels provide typical examples of the results we see across the region of the shell. Here we see the pixel sampled at $l,b=[0.0\deg,-24.96\deg]$ has a well constrained value of $\psi$, whereas the pixel at $l,b = [4.96\deg,-18.0\deg]$ does not. The pixel at $l,b = [4.96\deg,-18.0\deg]$ has an input PA difference very close to $0\deg$. In general we find that the value of $\psi$ is poorly constrained in regions where the input PA difference ($\text{PA}_\text{on}-\text{PA}_\text{off}$) is very close to $0\deg$, and in regions away from the bright background polarised emission of the Fermi bubbles.  The former case is a resultant behaviour of the Faraday screen model, which is better able to constrain values of $\psi$ which are significantly greater or smaller than $0\deg$. Additionally, in the lower panel of Figure~\ref{fig:post}, the blue distribution of the pixel at $l,b=[0.0\deg,-24.96\deg]$ has a dip around $\psi\approx0$. We find similar distributions for all pixels where the input PA difference is $~0\deg$. This can be seen in the lower panel of Figure~\ref{fig:faradaymap} as high values of the $1\sigma$ error. In this Figure we note that the outline of the shell region is visible due to the increased error inside the inner boundary of the shell, where $\text{PA}_\text{on}-\text{PA}_\text{off}\approx0\deg$. The uncertainties in $f$ and $c$ do follow similar trends to $\psi$, but overall the errors are much more uniform. Of note, the region around $l,b\approx[11\deg,-11\deg]$ corresponds to the highest error in $\psi$ and unusual values in all the output parameters. The region is also associated with very low polarised intensity from the WMAP and S-PASS observations, resulting in high uncertainty in that region.

\section{Discussion}\label{sec:results}

\subsection{Line of Sight Magnetic Field}\label{sec:magnitude}
The magnetic fields associated with \bub\ can be determined from its Faraday thickness. To compute this magnetic field strength we evaluate Equation~\ref{eqn:faradaydepth} along the line of sight:
\begin{equation}
	B_\parallel=\frac{\phi}{0.812\langle n_e\rangle L}
	\label{eqn:bsimple}
\end{equation}
where $B_\parallel$ and $\langle n_e\rangle$ are the line-of-sight of averages of the magnetic field and the electron density, respectively.

The problem that Equation~\ref{eqn:bsimple} presents is a ambiguity between $\phi$, $\langle n_e\rangle$, and $L$ when trying to determine $ B_\parallel $. We have already determined $ \phi $ using the Faraday screen model, which also sets the sign of the magnetic field along the line of sight; as $\langle n_e\rangle$, and $L$ are positive definite.

To constrain the path-length through the shell ($L$) we assume a simple spherical model. This model takes the path-length through the shell to be the chord between two concentric spheres. These spheres are co-centred and have an inner and outer radius such that they align with the centre, and the inner and outer bound of \bub, as observed in \hi. This path length is also a function of the distance to the centre of the shell. We provide a derivation for this path-length in Appendix~\ref{sec:path}.

To obtain the electron density information various different methods can be employed, each with its own advantages and disadvantages. We use the emission measure (EM) as determined by \halpha\ emission, following the procedure set out by \citet{Harvey-Smith2011a} and \citet{Gao2015}. Emission from \halpha\ can be observed in a diffuse manner over the entire sky; providing finely sampled spatial information on electron density. \bub\ is not clear in \halpha\ however, which means associating emission with the shell is difficult. Additionally, as a thermal process, the \halpha\ line is very broad. This makes constraining the emission to a particular velocity challenging; especially to velocities near the local standard of rest. If instead we use velocity integrated \halpha\ observations, we would of course be including emission from both in front of and behind our region of interest. In either case, making use of \halpha\ emission will at least provide an upper limit on the electron density. Computing an emission measure also requires information in addition to \halpha\ intensity. As such, the uncertainty associated with the electron density that is derived from an emission measure will be high. 

As an alternative, we did consider the electron density measured from the dispersion measure (DM) of pulsars behind \bub. To obtain the required DM data, we consult the ATNF Pulsar Catalogue \citep{Manchester2005}, and find thirteen pulsars within a $10\deg$ radius of the centre of the spherical region of \bub\footnote{Catalogue version: 1.57, Accessed 23$^\text{rd}$ of August 2017.}. Only three of these pulsars have distance estimates independent of the dispersion measure. Of these three, one sits on the other side of the Galactic centre, meaning that most of its dispersion measure likely comes from the Galactic central region. This makes any simplifying assumptions, such as a roughly constant $n_e$ unusable. The other two pulsars sit in front of the shell region. This means no usable electron density information was available from pulsar dispersion measures in the region of the shell. We also considered electron density information from other tracers such as \ion{S}{ii}, \ion{S}{iii}, and \ion{O}{ii}. However, we find from \citet{Shull2009} that intermediate velocity observations of ionised tracers, from both the \textit{Hubble Space Telescope} and \textit{Far Ultraviolet Spectroscopic Explorer}, do not probe near to the Galactic plane, and thus exclude the region around \bub. As each of these alternatives cannot provide us with with a value of $n_e$ in the region of \bub, we continue to use the EM to estimate this value.

\subsubsection{Emission Measure}
The emission measure is defined as the path integral of $n_e^2$ along the line of sight:
\begin{equation}
	\text{EM}\equiv\int_{0}^{L}n_e^2 dr
	\label{eqn:em}
\end{equation}
By constraining the limits of integration to within the path-length of the shell ($L$), we limit the EM obtained to be from the shell only. EM can be determined from observations of \halpha\ emission as follows \citep{Haffner1998a,Valls-Gabaud1998,Finkbeiner2003}:
\begin{equation}
	\text{EM} = 2.75T_{4}^{0.9}I_{\halpha}\exp{\left[2.44E(B-V)\right]}\,\text{pc\,cm$^{-6}$}
	\label{eqn:emhalpha}
\end{equation}
Again, if we constrain the measured $\halpha$ emission to within the shell, the resulting EM will be similarly constrained. Where $T_4$ is the thermal electron temperature in $10^4$\,K, $I_{\halpha}$ is the \halpha\ intensity in Rayleighs, and $E(B-V)$ is the colour excess \citep{Finkbeiner2003}. Note that to use this method, we require both an estimate of the electron temperature in the warm ionised medium (WIM) and measurements of the extinction from dust reddening. Here we assume a typical WIM temperature of $T_4\approx0.8\times10^4\,$K, following \citet{Gao2015} and \citet{Haffner1998a}. We obtain the colour excess from infrared dust measurements by \citet{Schlegel1998} and find a mean value within the region of the shell of $\sim0.17$. 

To find the electron density associated with the supershell we adopt the same formalism as \citet{Harvey-Smith2011a}. We allow the thermal electrons to be `clumped' along the line of sight. Outside of a clump we take $n_e=0$ and within a clump $n_e=n_{e,c}$. Using this information, Equation~\ref{eqn:em} can be solved for the electron density inside a clump:
\begin{equation}
	n_{e,c} = \sqrt{\frac{\text{EM}}{f_{e}L}}
	\label{eqn:neem}
\end{equation}
Note the addition of the thermal electron filling factor $f_{e}$. This term quantifies the line-of-sight distribution of thermal electrons. If the electrons are uniformly distributed then $f_{e}=1$; however, if the electrons are `clumped', or if the shell has an ionised layer, then $f_{e}<1$. That is, the electron clumps inhabit a column of $f_eL$ along the line of sight. From this, the average line-of-sight electron density is given by:
\begin{equation}
	\langle n_e\rangle = f_en_{e,c}
	\label{eqn:neav}
\end{equation}
Combining Equations \ref{eqn:bsimple}, \ref{eqn:neem}, and \ref{eqn:neav}  yields the line-of-sight magnetic field as:
\begin{equation}
B_{\parallel} = \frac{\phi}{0.812\sqrt{\text{EM}f_{e}L}}
\label{eqn:bem}
\end{equation}

We use the EM as obtained from velocity separated \halpha\ data, as it provides a more accurate estimate of the electron density within \bub. We obtain the kinematic data from the Wisconsin H-Alpha Mapper (WHAM) survey \citep{Haffner2003, Haffner2010}. The \halpha\ data from the WHAM kinematic survey will produce an upper limit on the electron density due to its broad line-width. We provide a map of EM in the Appendix in Figure~\ref{fig:em}. In the shell we find a mean EM of $12.6\,\text{pc\,cm$^{-6}$}$.

\begin{figure}
	\centering
	\includegraphics[width=\columnwidth]{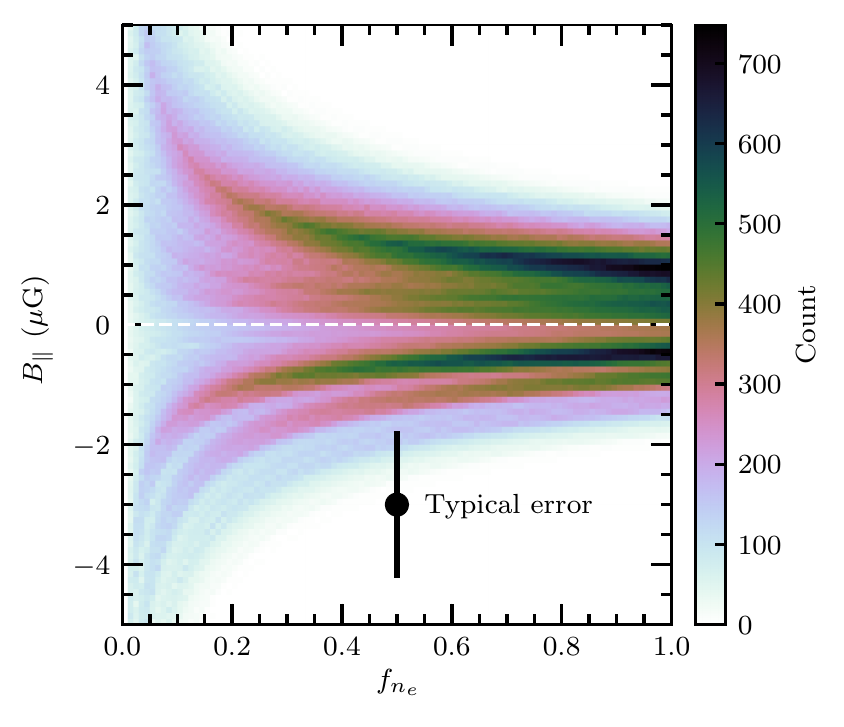}
	\caption{Distribution of the line-of-sight magnetic field ($B_\parallel$) as determined by the velocity separated \halpha, using Equation~\ref{eqn:bem}. We find distribution from the histogram of $B_\parallel$ between $\pm5\,\mu$G, with a bin-width of $0.1\,\mu$G, across the region of the shell, for each value of $f_{n_{e}}$. Note that the spread of values here does not necessarily correspond to error in $B_\parallel$, rather it shows the range of values found in the region in \bub. The `typical error' shown is the mean error in $B_\parallel$ when $f_{n_{e}}=0.5$.}
	\label{fig:halphaB}
\end{figure}

We now compute the line-of-sight magnetic field magnitude ($B_\parallel$) using Equation~\ref{eqn:bem}, taking a distance to the shell of 1.5\,kpc to evaluate the line-of-sight distance $L$ (see Section~\ref{sec:path}). We note here that the $B_\parallel$ we obtain is a lower limit, as the electron density derived from EM is an upper limit. In addition $f_{e}$ is not constrained, we therefore present $B_\parallel$ as a function of this factor. The spatial distribution of  $B_\parallel$ over the shell for each value of $f_{e}$ is given in Figure~\ref{fig:halphaB}. We also indicate in this Figure that the mean $1\sigma$ error in $B_\parallel$ is $\sim1.2\,\mu$G when $f_{e}=0.5$. As expected from Equation~\ref{eqn:bem}, the magnitude of $B_\parallel$ remains relatively constant as a function of $f_{e}$, until the filling factor becomes very small.

The value of $f_{e}$ is not well constrained, and as such a value is often assumed in the literature. A value of $f_{e}=1$, implying a uniform distribution of electrons, is unlikely; as are small values of $f_{e}$, since they imply large magnetic fields for a given Faraday depth measurement. We summarise some recent values of $f_{e}$ from the literature in Table~\ref{tab:fillingfactors}. \citet{Purcell2015} determined $f_{e}$ from an MCMC model fit to their data; as such, they constrain lower limit of $f_{e}=0.24$ and a mean value of $f_{e}=0.3$. They note, however, that a value of around 0.5 provided a better match to dispersion measure data from pulsars. \protect\citet{Kaczmarek2017} adopted their value of 0.5 following \protect\citet{McClure-Griffiths2010}. We note that some of these values are not directly derived, but rather chosen based on previous studies. Considering these values, and the range over which our derived field strength remains relatively constant, we will now adopt a value of $f_{e}=0.5$ for further analysis.

\begin{table}
	\caption{A summary of recent literature values of $f_{e}$. Recall $f_e$ is the filling factor of thermal electrons. (HVC - High velocity cloud.) } 
	\begin{tabular}{c l l}
		\hline
		$f_{e}$			&	Phenomena				&		Work									\\   
		\hline														
		$0.04\pm0.01$	 	& Mid-plane WIM  			& 		\protect\citet{Gaensler2008} 			\\
        $\sim0.3$ 			& Off-plane WIM				&		\protect\citet{Gaensler2008} 			\\
        0.5					& HVC						&   	\protect\citet{McClure-Griffiths2010}	\\
		0.1					& \hii\ regions				&   	\protect\citet{Harvey-Smith2011a} 		\\
		$0.4$ - $1$			& W4 Superbubble			&   	\protect\citet{Gao2015} 				\\
		$\geq0.24$  		& Gum nebula				&   	\protect\citet{Purcell2015}				\\
		0.5   				& Magellanic Bridge			&   	\protect\citet{Kaczmarek2017}   		\\
		\hline
	\end{tabular}
	\label{tab:fillingfactors}
\end{table}

\begin{figure*}
	\centering
	\begin{subfigure}{0.49\textwidth}
			\includegraphics[width=\columnwidth]{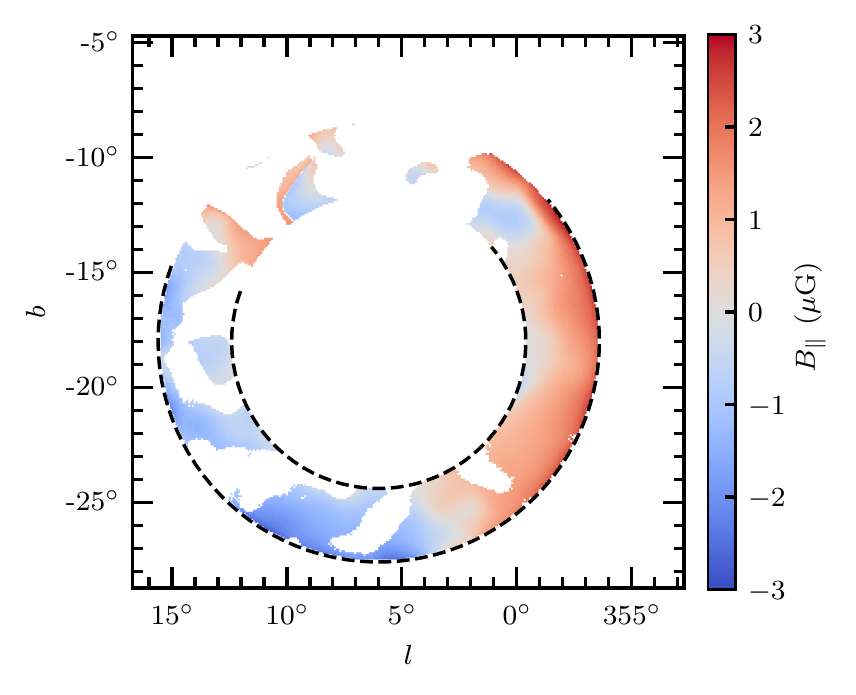}
		\caption{}
		\label{fig:bmap}
	\end{subfigure}
	\begin{subfigure}{0.49\textwidth}
		\includegraphics[width=\columnwidth]{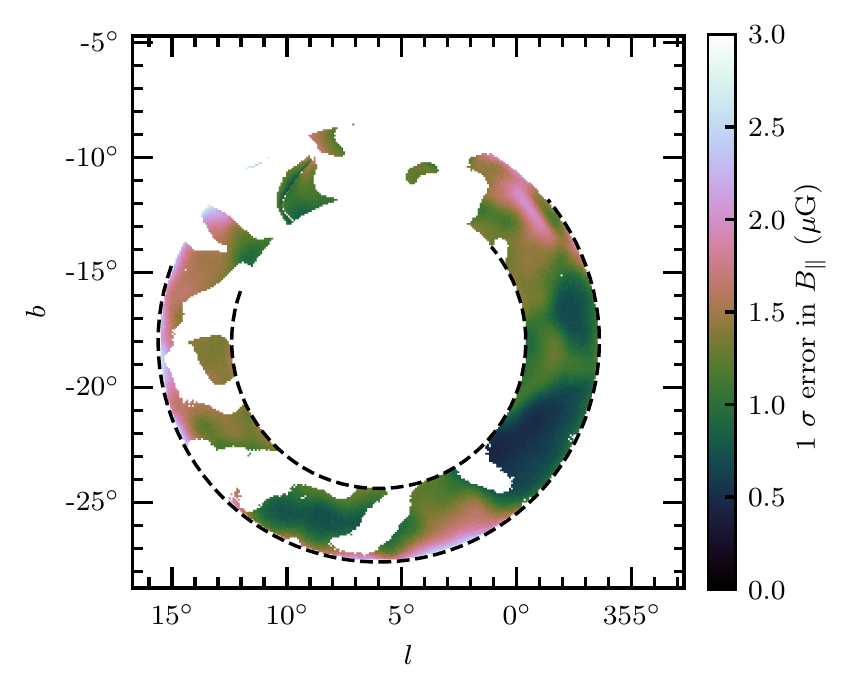}
		\caption{}
		\label{fig:bmaperr}
	\end{subfigure}
	\caption{\subref{fig:bmap} The line-of-sight magnetic field map through \bub. Here $B_\parallel$ is determined from Equation~\ref{eqn:bem} using EM from velocity separated \halpha, taking $f_{e}=0.5$ and a distance to the shell of 1.5\,kpc. We provide a map of EM in Figure~\ref{fig:em}. Note that some extreme values of the magnetic field occur towards the outer boundary of the shell, and arise due to the path-length $L$ becoming very small at the limb. \subref{fig:bmaperr} The map of uncertainty in the line-of-sight magnetic field. }

\end{figure*}

\subsection{Line of Sight Magnetic Field Structure}\label{sec:structure}
We provide a map of the line-of-sight magnetic field structure in Figure~\ref{fig:bmap} taking $f_{e}=0.5$ and a distance of 1.5\,kpc. The uncertainty of the values in this map is provided in Figure~\ref{fig:bmaperr}. The structure presented in the map of the line-of-sight B-field appears to follow the general trend of away from the observer (negative) in the bottom-left of the shell; transitioning to toward the observer in the top-right of the shell. This structure is, upon simple consideration, consistent with a field that is azimuthally wrapped around the surface of the shell.

\citet{Stil2009} analysed the role of magnetic fields in expanding superbubbles through MHD simulations. Their work also provided simulations of RM signatures from superbubbles expanding from the Galactic plane with a magnetic field parallel to the plane. This provides two examples to compare with our Faraday depth map (Figure~\ref{fig:faradaymask}): an observation looking perpendicular to the Galactic magnetic field, and an observation looking parallel. In the former case, the strongest RM values are found in lines of sight through the cavity of the simulated bubble. Most notably, the sign of the RM in the shell of the bubble reverses along the line that goes through the centre of the bubble, parallel to the external field direction. In the case of observing parallel to the Galactic magnetic field the RM values are greatly amplified overall, and the wall of the bubble exhibits the strongest RM signature. Between these two scenarios, our Faraday thickness map is in better agreement with a shell which has expanded into a field perpendicular to the line of sight as we see a very clear sign change in the shell from lower left to upper right. It is likely, however, that we observe this shell at some angle between the two cases described by \citet{Stil2009}. This is an area that could be further probed by simulation work. If the geometry and orientation of the shell can be well determined, and therefore the total magnetic field structure, this information could provide insight into the Galactic field into which the shell expanded.

\subsection{Implications}\label{sec:implications}

\subsubsection{Dynamical Role of B-fields in \bub}
The dynamical importance of magnetic fields in an ionised medium can be quantified by the plasma beta ($\beta_\text{th}$):
\begin{equation}
	\beta_\text{th} = \frac{P_\text{th}}{P_\text{mag}}
	\label{eqn:plamabeta}
\end{equation}
which is the ratio of the thermal pressure ($P_\text{th}$) to the magnetic pressure ($P_\text{mag}$). In this analysis the plasma beta quantifies the dynamical role of the magnetic fields in the shell itself.

We will assume that the shell contains a mixture of a warm neutral medium (WNM) and WIM as described by \citet{Heiles2012} and that the value of $B_\parallel$ remains the same in both phases. This is motivated by the observations of \hi\ and \halpha\ in this region. We are therefore also assuming that the measured \hi\ emission, \halpha\ emission, and Faraday depth, all arise from the same location. The thermal pressure is therefore the sum of the partial pressures of the ionised ($ P_{\text{th},i} $) and neutral media ($ P_{\text{th},n} $):
\begin{equation}
	P_\text{th} = \langle P_{\text{th},i} \rangle + \langle P_{\text{th},n} \rangle
	\label{eqn:pth}
\end{equation}
where the terms in angular parentheses refer to the line-of-sight averages of those values. The partial pressures are given by:
\begin{equation}
	\begin{aligned}
	\langle P_{\text{th},i} \rangle &= 2\langle n_e\rangle k T_i = 2f_en_ekT_i\\
	\langle P_{\text{th},n} \rangle &= \langle n_H \rangle k T_n
	\end{aligned}
	\label{eqn:pthin}
\end{equation}
where $n_e$ is the electron number density derived from EM, $n_H$ is the neutral hydrogen number density, $k$ is Boltzmann's constant, and $T_i$ and $T_n$ are the temperatures of the ionised and neutral phase, respectively. We take values of $T_i=8000\,$K for the ionised medium and $T_n=6000\,$K for the neutral medium. We obtain the number density of neutral hydrogen from the column density derived by \citet{Moss2012} from GASS \hi. They find a mean column density of $N_{H,\text{av}}\sim2\times 10^{20}\,\text{cm}^{-2}$ in the shell of \bub. From this, we find the line-of-sight averaged number density of \hi\ in the shell from:
\begin{equation}
\langle n_H \rangle = \frac{N_{H,\text{av}}}{L}
\label{eqn:numberdens}
\end{equation}
where $L$ is the path-length through the shell (see Appendix~\ref{sec:path}). Note that $L$ varies across the projected area of the shell, therefore we also obtain a spatially varying value of $n_H$. In this region we find mean values of $\langle n_e\rangle$ and $\langle n_H \rangle$ of 0.15\,cm$^{-3}$ and 0.28\,cm$^{-3}$, respectively. Recall, however, that our value of $\langle n_e\rangle$, as derived from EM, is an upper limit.

\begin{figure*}
	\centering
	\includegraphics[width=2\columnwidth]{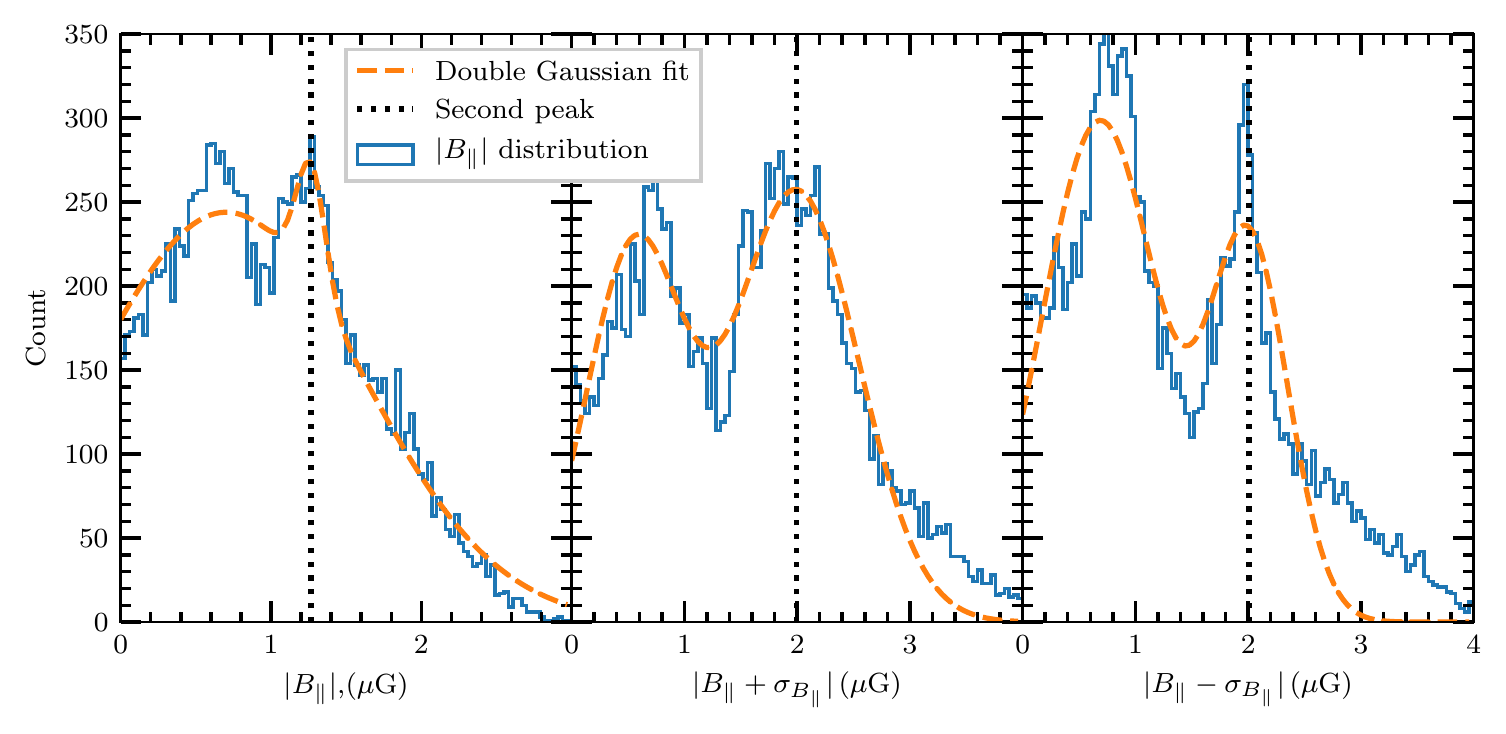}
	\caption{The distribution of the absolute value line-of-sight magnetic field ($|B_\parallel|$) across \bub. The left panel shows the distribution of $|B_\parallel|$ taking $f_{e}=0.5$ and a distance to the shell of 1.5\,kpc (i.e. the distribution of values from Figure~\ref{fig:bmap}). The middle and right-hand panels show the distribution of $|B_\parallel|$ plus and minus the error in $B_\parallel$ (as shown in Figure~\ref{fig:bmaperr}), respectively. Each distribution has been fit with a double Gaussian (i.e. the sum of two Gaussians), shown in orange. The centre of the second fitted peak is shown by the black dashed line. The locations of the second peak for the left, middle, and right panels are 1.27\,$\mu$G, 1.99\,$\mu$G, and 2.00\,$\mu$G, respectively.}
	\label{fig:bdists}
\end{figure*}

The magnetic pressure in the shell in given by:
\begin{equation}
P_\text{mag} = \frac{B^2_\text{tot}}{8\pi}
\label{eqn:pmag}
\end{equation}
where $B_\text{tot}$ is the total local magnetic field. Our observations have provided us with the line-of-sight field, however. If we consider the case of an azimuthally wrapped magnetic field within the shell, we expect the line-of-sight field to have a distribution across the shell. That is, maximum when the total field is aligned with the line of sight, and null when the field is perpendicular. Additionally, as our model for the path length through the shell has a hard boundary, the values of this length become very small towards the edge of the shell and thus resulting in large $|B_\parallel|$. Overall what we expect from the distribution of $|B_\parallel|$ over the shell is a smooth peak near small values of $|B_\parallel|$, a peak at the value which corresponds to the field being aligned with the line of sight, and a tail of more extreme values of $|B_\parallel|$. We find a similar distribution to this across \bub, as shown in the left panel of Figure~\ref{fig:bdists}. This distribution is bi-modal; we interpret the first peak to correspond to regions where the total field is close to perpendicular to the line of sight and the second peak to where the total field is close to parallel. To find the locations of these peaks we fit a double Gaussian (i.e. the sum of two Gaussians) to the distribution. To obtain the uncertainty of the peak value we perform the same analysis of the distribution of $|B_\parallel+\sigma_{B_\parallel}|$ and $|B_\parallel-\sigma_{B_\parallel}|$, as shown in the middle and right-hand panels of Figure~\ref{fig:bdists}. From this we obtain a value of the second peak of $|B_\parallel|_\text{peak} = 2.0\substack{+0.01 \\ -0.7}\,\mu$G. The error range given here includes uncertainties arising from our best-fit $\phi$ value, $\langle n_e\rangle$ estimate, and our model $L$. We now assume that $B_\text{tot}=|B_\parallel|_\text{peak}$, since this value is likely associated with the total field being aligned with the line of sight in the case of a coherent total field. Recall that as $B_\parallel$ depends on the value we found for EM, which is an upper limit, the value of $B_\parallel$ and our approximation of $B_\text{tot}$ are therefore lower limits. We note it would be possible for a stronger magnetic field to be obtained from our modelled Faraday depth if the electron density or $f_e$ were demonstrated to be smaller than our current estimates. Additional data would be required, however, to motivate a different estimation.

We compute the mean plasma beta across \bub\ using both the ionised and neutral partial pressures as described above. From this we obtain a plasma beta of $\beta_\text{th} = 4 \substack{+11 \\ -2}$. We note that this value has a high variance, which is due to the sensitivity of $\beta_\text{th}$ to small values of $B_\text{tot}$. Additionally, the value of $\beta_\text{th}$ is an upper limit only, as $B_\text{tot}$ was a lower limit and $\langle n_e\rangle$ was an upper limit. The errors given here include uncertainties in $\langle n_e\rangle$, $\langle n_H\rangle$, and $B_\text{tot}$. This value of $\beta_\text{th}$ implies that magnetic field pressures in \bub\ are dominated by thermal pressures in the region of the shell. Due to the large uncertainties involved it is hard to draw further conclusions regarding the dynamical role of magnetic fields in \bub. Such analysis would also require additional information about thermal pressures in the cavity of the shell \citep{Ferriere1991}, which is beyond the scope of this work. What is of note, however, is that despite how relatively weak the magnetic fields in \bub\ are, this technique has allowed their detection.

\subsubsection{Comparison to other results}\label{sec:compare}
Two structures of similar origin to \bub\ have recently had measurements of their associated B-fields. \citet{Gao2015} analyse the magnetic fields associated with the W4 superbubble, and \citet{Purcell2015} study the Gum Nebula. \citet{Gao2015} find strong magnetic fields in association with the W4 superbubble ($B_\parallel=-5.0\,\mu\text{G}/\sqrt{f_{e}}\pm10\%$, $B_\text{tot}>12\,\mu$G), which generally dominate the thermal pressures in the Western wall of the shell. They also find that towards the high-latitude region of W4 the magnetic fields weaken; making the magnetic and thermal pressures comparable. In the Gum Nebula, \citet{Purcell2015} find a total magnetic field strength of $B_\text{tot}=3.9\substack{+4.9 \\ -2.2}\,\mu$G. From this they compute a plasma beta $\beta_\text{th}=4.8$, which is relatively high, meaning that the thermal pressures dominate the region, similar to our findings for \bub.

\begin{figure*}
	\centering
	\includegraphics[width=2\columnwidth]{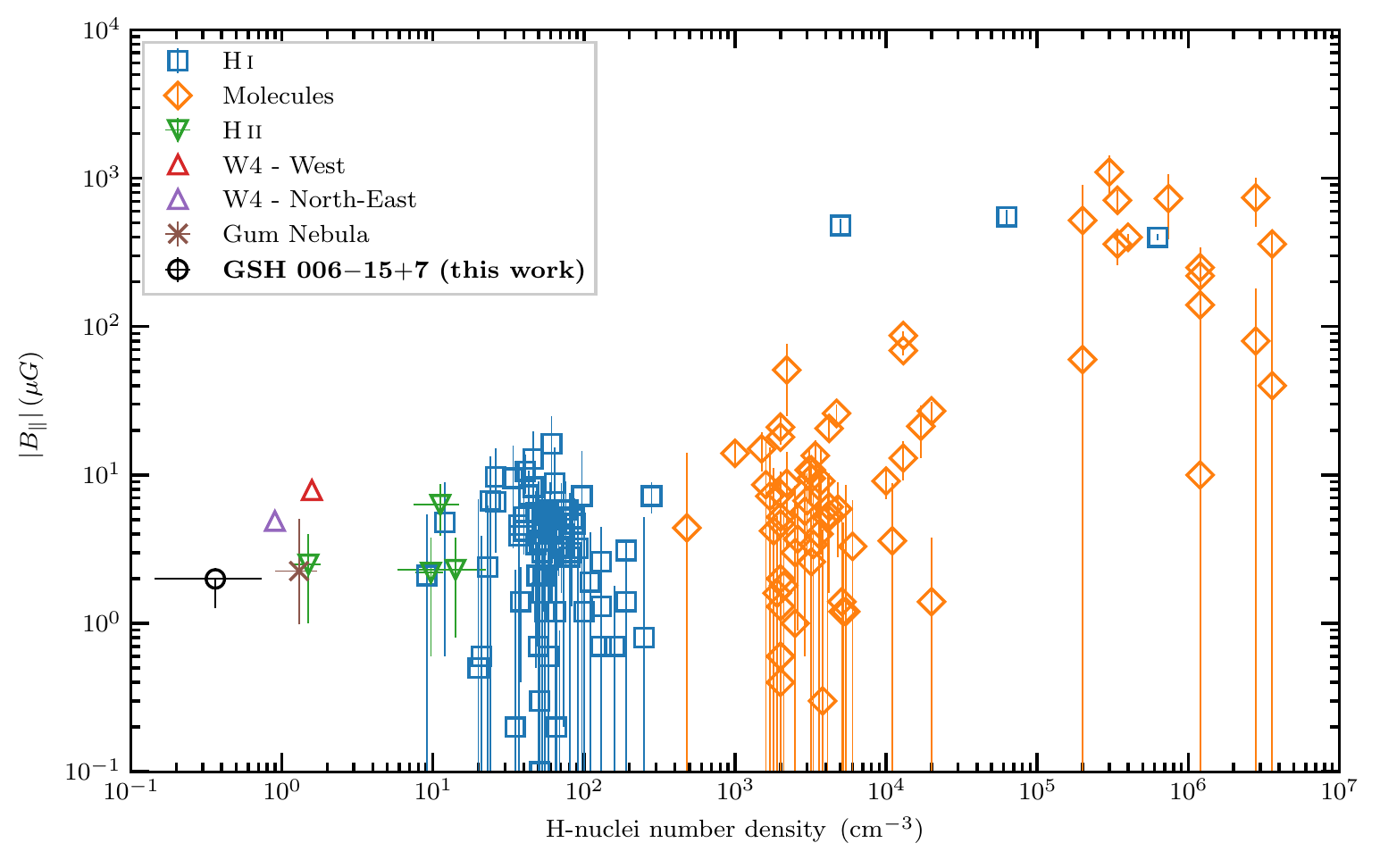}
	\caption{Line-of-sight magnetic field amplitude against number density in the associated ISM. The \hi\ and molecular data were originally compared by \protect\citet{Crutcher2010}, and the \hii\ data were added by \citet{Harvey-Smith2011a}. We make the addition of \bub\ (this work) using the RMS value of the field, as well as the W4 superbubble \protect\citep{Gao2015} and the Gum Nebula~\protect\citep{Purcell2015}. Note that we have computed the line-of-sight B-field strength in the Gum Nebula from the total field strength by $B_\parallel\sim B_\text{tot}/\sqrt{3}$.}
	\label{fig:compare}
\end{figure*}

To place these values in a broader context, we compare these results to magnetic fields found in molecular and \hi\ clouds by \citet{Crutcher2010} and in \hii\ regions by \citet{Harvey-Smith2011a}. In Figure~\ref{fig:compare} we add this work on \bub, as well as the W4 and Gum Nebula results, to the comparison of density against magnetic field strength. We find the H-nuclei number density ($ n_{\text{H-nuclei}} $) in \bub\ using $n_{\text{H-nuclei}}=\langle n_H\rangle + \langle n_p\rangle = \langle n_H\rangle + f_en_e$, where $n_p$ is the number density of protons in the ionised phase. We note that the largest objects appear in a cluster together in the lowest density region of this Figure. The \hii\ region that appears along with \bub, W4, and the Gum Nebula, is Sivian 3; which is the largest \hii\ region analysed by \citet{Harvey-Smith2011a}. We find that \bub\ has comparable magnetic fields amongst these objects, but is slightly weaker and occupies a lower density regime. The resulting plasma beta for this object is therefore indicative that thermal pressures dominate the shell.

\begin{figure}
	\centering
	\includegraphics[width=\columnwidth]{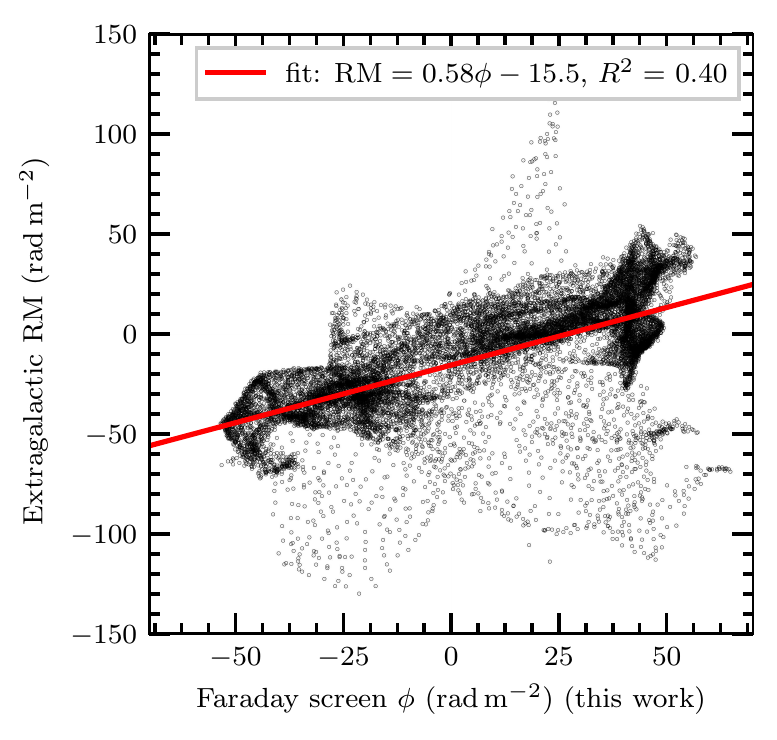}
	\caption{The comparison of extragalactic RMs from \citet{Oppermann2012} with the Faraday depth found from the Faraday screen model in the region of \bub. The solid red line corresponds to a least-squares linear fit to the data.}
	\label{fig:exgal}
\end{figure}

Finally, we compare the Faraday thickness derived from the Faraday screen model with RMs from extragalactic sources. If \bub\ is the dominant Faraday component along the line-of-sight, we expect to see evidence of it in these data. We obtain the all-sky map of extragalactic RMs from \citet{Oppermann2015}. We find a correlation between these extragalactic RMs and the Faraday thicknesses from the screen model, as shown in Figure~\ref{fig:exgal}. There is significant scatter present in this correlation; which is as expected as extragalactic RMs probe the entire line-of-sight through the Galaxy, and thus multiple Faraday components. We fit a linear model to these data, and find that $\text{RM}\sim0.6\phi-15.5$, with a coefficient of determination $R^2=0.4$. The physical reason for the slope correlation is not obvious, as a factor of $1/2$ is usually expected for regions of mixed emission and rotation \citep[e.g.][]{Sokoloff1998}.

\section{Summary and Conclusions}\label{sec:summary}
We found the polarised signature at 2.3\,GHz of Galactic supershell \bub\ in S-PASS. The morphological correlation indicates a direct detection of the MIM in association with this Galactic supershell. The `shadow' of \bub\ is most obvious in the Stokes $U$ images. While there are signatures of the shell in polarised intensity and in Stokes $Q$, they are not as obvious. This highlights the importance of investigating multiple polarisation modes when searching for polarised features.

We have provided a method of obtaining a Faraday depth map from two single-frequency observations; modifying the approach for modelling a Faraday screen. To estimate the background synchrotron emission of this object we used high-frequency polarisation observations from WMAP K-band scaled to S-band. This has allowed us to obtain Faraday rotation information with fewer assumptions than just using single-frequency observations. This method can be applied wherever a bright polarised background source illuminates purely Faraday-rotating foreground object. The source itself can be extended and complex in structure; so long as the Faraday rotation in the screen remains between $-90\deg<\psi<+90\deg$, the Faraday depth can be successfully recovered. This condition can be verified through inspection of a single-frequency polarisation-angle map. In Section~\ref{sec:bestfit} we describe our best-fit procedure, which we find to be robust and parallelisable on multi-core computers. 

From this Faraday depth map we determine the line-of-sight magnetic field and structure in association with the Galactic supershell \bub. We derive the line-of-sight field in Section~\ref{sec:magnitude}, and discuss the structure in Section~\ref{sec:structure}. We find a peak line-of-sight field strength of $|B_\parallel|_\text{peak} = 2.0\substack{+0.01 \\ -0.7}\,\mu$G. From these results we have gained insight into the dynamical role of the magnetic fields associated with \bub. In the region of the shell we find that the magnetic pressures are likely dominated by thermal pressures. The line-of-sight magnetic field structure indicates that the Galactic magnetic field has a significant component perpendicular to the line of sight in the region of \bub.

We find that the line-of-sight field strength is comparable to similarly sized objects with similar densities. This indicates that by using diffuse polarisation observations we are able to probe the magnetic fields in low-density regimes of the ISM. The method we have developed has a number of advantages in comparison to observations of point-source RMs. Most relevant to this work is that our method allows the Faraday thickness of a single, extended object to be constrained, even though it lies close to the Galactic plane. 

As this object was illuminated by an extended, bright polarised source in the background, we are still able to detect Faraday rotation and assume a Faraday screen interaction. Point-source RMs probe all Faraday depths along the line of sight, which becomes very complex near the Galactic plane. Secondly, the magnetic fields detected here are weak relative to most of the previous measurements from other ISM sources reported in the literature. As such, we have shown this method to be useful in the detection of weak magnetic fields in large and diffuse areas.

\section*{Acknowledgements}
The authors would like to acknowledge the significant contributions of the reviewer, Katia Ferriere, whose comments have made a substantial improvement to this work.

A.T. acknowledges the support of the Australian Government Research Training Program (RTP) Scholarship. N. M. M.-G. acknowledges the support of the Australian Research Council through grant FT150100024. C.F.~gratefully acknowledges funding provided by the Australian Research Council's Discovery Projects (grants~DP150104329 and~DP170100603) and the ANU Futures Scheme. The Dunlap Institute is funded through an endowment established by the David Dunlap family and the University of Toronto. B.M.G. acknowledges the support of the Natural Sciences and Engineering Research Council of Canada (NSERC) through grant RGPIN-2015-05948, and of the Canada Research Chairs program. The parallel data analysis presented in this work used high performance computing resources provided by the Australian National Computational Infrastructure (grant~ek9), and the Pawsey Supercomputing Centre with funding from the Australian Government and the Government of Western Australia, in the framework of the National Computational Merit Allocation Scheme and the ANU Allocation Scheme. The Parkes Radio Telescope is part of the Australia Telescope National Facility, which is funded by the Commonwealth of Australia for operation as a National Facility managed by CSIRO. This work has been carried out in the framework of the S-band Polarisation All Sky Survey (S-PASS) collaboration.

The authors acknowledge the use of the following archival data: 
\begin{itemize}
	\item The Legacy Archive for Microwave Background Data Analysis (LAMBDA), part of the High Energy Astrophysics Science Archive Centre (HEASARC). HEASARC/LAMBDA is a service of the Astrophysics Science Division at the NASA Goddard Space Flight Centre.
	\item The Southern H-Alpha Sky Survey Atlas (SHASSA), which is supported by the National Science Foundation.
	\item The Wisconsin \halpha\ Mapper and its \halpha\ Sky Survey, funded primarily by the National Science Foundation. The facility was designed and built with the help of the University of Wisconsin Graduate School, Physical Sciences Lab, and Space Astronomy Lab. NOAO staff at Kitt Peak and Cerro Tololo provided on-site support for its remote operation.
\end{itemize}




\bibliographystyle{mnras}
\bibliography{final} 

\begin{thebibliography}{}
\makeatletter
\relax
\def\mn@urlcharsother{\let\do\@makeother \do\$\do\&\do\#\do\^\do\_\do\%\do\~}
\def\mn@doi{\begingroup\mn@urlcharsother \@ifnextchar [ {\mn@doi@}
  {\mn@doi@[]}}
\def\mn@doi@[#1]#2{\def\@tempa{#1}\ifx\@tempa\@empty \href
  {http://dx.doi.org/#2} {doi:#2}\else \href {http://dx.doi.org/#2} {#1}\fi
  \endgroup}
\def\mn@eprint#1#2{\mn@eprint@#1:#2::\@nil}
\def\mn@eprint@arXiv#1{\href {http://arxiv.org/abs/#1} {{\tt arXiv:#1}}}
\def\mn@eprint@dblp#1{\href {http://dblp.uni-trier.de/rec/bibtex/#1.xml}
  {dblp:#1}}
\def\mn@eprint@#1:#2:#3:#4\@nil{\def\@tempa {#1}\def\@tempb {#2}\def\@tempc
  {#3}\ifx \@tempc \@empty \let \@tempc \@tempb \let \@tempb \@tempa \fi \ifx
  \@tempb \@empty \def\@tempb {arXiv}\fi \@ifundefined
  {mn@eprint@\@tempb}{\@tempb:\@tempc}{\expandafter \expandafter \csname
  mn@eprint@\@tempb\endcsname \expandafter{\@tempc}}}

\bibitem[\protect\citeauthoryear{Ackermann et~al.,}{Ackermann
  et~al.}{2014}]{Ackermann2014}
Ackermann M.,  et~al., 2014, \mn@doi [\apj] {10.1088/0004-637X/793/1/64}, 793,
  64

\bibitem[\protect\citeauthoryear{{Ben Bekhti} et~al.,}{{Ben Bekhti}
  et~al.}{2016}]{BenBekhti2016}
{Ben Bekhti} N.,  et~al., 2016, \mn@doi [\aap] {10.1051/0004-6361/201629178},
  594, A116

\bibitem[\protect\citeauthoryear{Bennett et~al.,}{Bennett
  et~al.}{2013}]{Bennett2013}
Bennett C.~L.,  et~al., 2013, \mn@doi [\apjs] {10.1088/0067-0049/208/2/20},
  208, 20

\bibitem[\protect\citeauthoryear{Boumis, Dickinson, Meaburn, Goudis,
  Christopoulou, Lopez, Bryce  \& Redman}{Boumis et~al.}{2001}]{Boumis2001}
Boumis P.,  Dickinson C.,  Meaburn J.,  Goudis C.~D.,  Christopoulou P.~E.,
  Lopez J.~A.,  Bryce M.,   Redman M.~P.,  2001, \mn@doi [\mnras]
  {10.1046/j.1365-8711.2001.03950.x}, 320, 61

\bibitem[\protect\citeauthoryear{Brentjens \& de Bruyn}{Brentjens \&
  de~Bruyn}{2005}]{Brentjens2005}
Brentjens M.~A.,  de Bruyn A.~G.,  2005, \mn@doi [\aap]
  {10.1051/0004-6361:20052990}, 441, 1217

\bibitem[\protect\citeauthoryear{Burn}{Burn}{1966}]{Burn1966}
Burn B.~J.,  1966, \mn@doi [\mnras] {10.1093/mnras/133.1.67}, 133, 67

\bibitem[\protect\citeauthoryear{Calabretta, Staveley-Smith  \&
  Barnes}{Calabretta et~al.}{2014}]{Calabretta2014}
Calabretta M.~R.,  Staveley-Smith L.,   Barnes D.~G.,  2014, \mn@doi [\pasa]
  {10.1017/pasa.2013.36}, 31

\bibitem[\protect\citeauthoryear{Carretti et~al.,}{Carretti
  et~al.}{2013}]{Carretti2013}
Carretti E.,  et~al., 2013, \mn@doi [\nat] {10.1038/nature11734}, 493

\bibitem[\protect\citeauthoryear{Crutcher, Wandelt, Heiles, Falgarone  \&
  Troland}{Crutcher et~al.}{2010}]{Crutcher2010}
Crutcher R.~M.,  Wandelt B.,  Heiles C.,  Falgarone E.,   Troland T.~H.,  2010,
  \mn@doi [\apj] {10.1088/0004-637X/725/1/466}, 725, 466

\bibitem[\protect\citeauthoryear{Duncan, Haynes, Jones  \& Stewart}{Duncan
  et~al.}{1997}]{Duncan1997}
Duncan A.~R.,  Haynes R.~F.,  Jones K.~L.,   Stewart R.~T.,  1997, \mn@doi
  [\mnras] {10.1093/mnras/291.2.279}, 291, 279

\bibitem[\protect\citeauthoryear{Ehlerov{\'{a}} \& Palou{\v{s}}}{Ehlerov{\'{a}}
  \& Palou{\v{s}}}{2005}]{Ehlerova2005}
Ehlerov{\'{a}} S.,  Palou{\v{s}} J.,  2005, \mn@doi [\aap]
  {10.1051/0004-6361:20034389}, 437, 101

\bibitem[\protect\citeauthoryear{Ehlerov{\'{a}} \& Palou{\v{s}}}{Ehlerov{\'{a}}
  \& Palou{\v{s}}}{2013}]{Ehlerova2013}
Ehlerov{\'{a}} S.,  Palou{\v{s}} J.,  2013, \mn@doi [\aap]
  {10.1051/0004-6361/201220341}, 550, A23

\bibitem[\protect\citeauthoryear{Ferri{\`{e}}re}{Ferri{\`{e}}re}{2001}]{Ferriere2001}
Ferri{\`{e}}re K.~M.,  2001, \mn@doi [Reviews of Modern Physics]
  {10.1103/RevModPhys.73.1031}, 73, 1031

\bibitem[\protect\citeauthoryear{Ferriere, {Mac Low}  \& Zweibel}{Ferriere
  et~al.}{1991}]{Ferriere1991}
Ferriere K.~M.,  {Mac Low} M.-M.,   Zweibel E.~G.,  1991, \mn@doi [\apj]
  {10.1086/170185}, 375, 239

\bibitem[\protect\citeauthoryear{Finkbeiner}{Finkbeiner}{2003}]{Finkbeiner2003}
Finkbeiner D.~P.,  2003, \mn@doi [\apjs] {10.1086/374411}, 146, 407

\bibitem[\protect\citeauthoryear{Gaensler, Madsen, Chatterjee  \& Mao}{Gaensler
  et~al.}{2008}]{Gaensler2008}
Gaensler B.~M.,  Madsen G.~J.,  Chatterjee S.,   Mao S.~A.,  2008, \mn@doi
  [\pasa] {10.1071/AS08004}, 25, 184

\bibitem[\protect\citeauthoryear{Gao et~al.,}{Gao et~al.}{2010}]{Gao2010}
Gao X.~Y.,  et~al., 2010, \mn@doi [\aap] {10.1051/0004-6361/200913793}, 515,
  A64

\bibitem[\protect\citeauthoryear{Gao, Reich, Reich, Han  \& Kothes}{Gao
  et~al.}{2015}]{Gao2015}
Gao X.~Y.,  Reich W.,  Reich P.,  Han J.~L.,   Kothes R.,  2015, \mn@doi [\aap]
  {10.1051/0004-6361/201424952}, 578, A24

\bibitem[\protect\citeauthoryear{Haffner, Reynolds  \& Tufte}{Haffner
  et~al.}{1998}]{Haffner1998a}
Haffner L.~M.,  Reynolds R.~J.,   Tufte S.~L.,  1998, \mn@doi [\apj]
  {10.1086/311449}, 501, L83

\bibitem[\protect\citeauthoryear{Haffner, Reynolds, Tufte, Madsen, Jaehnig  \&
  Percival}{Haffner et~al.}{2003}]{Haffner2003}
Haffner L.~M.,  Reynolds R.~J.,  Tufte S.~L.,  Madsen G.~J.,  Jaehnig K.~P.,
  Percival J.~W.,  2003, \mn@doi [\apjs] {10.1086/378850}, 149, 405

\bibitem[\protect\citeauthoryear{Haffner et~al.,}{Haffner
  et~al.}{2010}]{Haffner2010}
Haffner L.~M.,  et~al., 2010, The Dynamic Interstellar Medium: A Celebration of
  the Canadian Galactic Plane Survey. Proceedings of a conference held at the
  Naramata Centre, 438, 388

\bibitem[\protect\citeauthoryear{Han}{Han}{2017}]{Han2017}
Han J.,  2017, \mn@doi [\araa] {10.1146/annurev-astro-091916-055221}, 55, 111

\bibitem[\protect\citeauthoryear{Harvey-Smith, Madsen  \&
  Gaensler}{Harvey-Smith et~al.}{2011}]{Harvey-Smith2011a}
Harvey-Smith L.,  Madsen G.~J.,   Gaensler B.~M.,  2011, \mn@doi [\apj]
  {10.1088/0004-637X/736/2/83}, 736, 83

\bibitem[\protect\citeauthoryear{Heiles}{Heiles}{1979}]{Heiles1979}
Heiles C.,  1979, \mn@doi [\apj] {10.1086/156986}, 229, 533

\bibitem[\protect\citeauthoryear{Heiles}{Heiles}{1984}]{Heiles1984}
Heiles C.,  1984, \mn@doi [ApJs] {10.1086/190970}, 55, 585

\bibitem[\protect\citeauthoryear{Heiles \& Haverkorn}{Heiles \&
  Haverkorn}{2012}]{Heiles2012}
Heiles C.,  Haverkorn M.,  2012, \mn@doi [\ssr] {10.1007/s11214-012-9866-4},
  166, 293

\bibitem[\protect\citeauthoryear{Heiles, Haffner  \& Reynolds}{Heiles
  et~al.}{1999}]{Heiles1999}
Heiles C.,  Haffner L.~M.,   Reynolds R.~J.,  1999, \mn@doi [New Perspectives
  on the Interstellar Medium] {1999ASPC..168..211H}, 168, 211

\bibitem[\protect\citeauthoryear{Hu}{Hu}{1981}]{Hu1981a}
Hu E.~M.,  1981, \mn@doi [\apj] {10.1086/159135}, 248, 119

\bibitem[\protect\citeauthoryear{Jo, Min, Seon, Edelstein  \& Han}{Jo
  et~al.}{2011}]{Jo2011}
Jo Y.-S.,  Min K.-W.,  Seon K.-I.,  Edelstein J.,   Han W.,  2011, \mn@doi
  [\apj] {10.1088/0004-637X/738/1/91}, 738, 91

\bibitem[\protect\citeauthoryear{Jo, Min  \& Seon}{Jo et~al.}{2015}]{Jo2015}
Jo Y.-S.,  Min K.-W.,   Seon K.-I.,  2015, \mn@doi [\apj]
  {10.1088/0004-637X/807/1/68}, 807, 68

\bibitem[\protect\citeauthoryear{Kaczmarek, Purcell, Gaensler,
  McClure-Griffiths  \& Stevens}{Kaczmarek et~al.}{2017}]{Kaczmarek2017}
Kaczmarek J.~F.,  Purcell C.~R.,  Gaensler B.~M.,  McClure-Griffiths N.~M.,
  Stevens J.,  2017, \mn@doi [\mnras] {10.1093/mnras/stx206}, 467, stx206

\bibitem[\protect\citeauthoryear{Kalberla \& Haud}{Kalberla \&
  Haud}{2015}]{Kalberla2015}
Kalberla P. M.~W.,  Haud U.,  2015, \mn@doi [\aap]
  {10.1051/0004-6361/201525859}, 578, A78

\bibitem[\protect\citeauthoryear{Kalberla, McClure-Griffiths  \& Kerp}{Kalberla
  et~al.}{2009}]{Kalberla2009}
Kalberla P. M.~W.,  McClure-Griffiths N.~M.,   Kerp J.,  2009, Proceedings of
  Panoramic Radio Astronomy

\bibitem[\protect\citeauthoryear{Koo, Heiles  \& Reach}{Koo
  et~al.}{1992}]{Koo1992}
Koo B.-C.,  Heiles C.,   Reach W.,  1992, \mn@doi [ApJ] {10.1086/171264}, 390,
  108

\bibitem[\protect\citeauthoryear{Maciejewski, Murphy, Lockman  \&
  Savage}{Maciejewski et~al.}{1996}]{Maciejewski1996}
Maciejewski W.,  Murphy E.,  Lockman F.,   Savage B.,  1996, \mn@doi [ApJ]
  {10.1086/177774}, 469, 238

\bibitem[\protect\citeauthoryear{Manchester, Hobbs, Teoh  \& Hobbs}{Manchester
  et~al.}{2005}]{Manchester2005}
Manchester R.~N.,  Hobbs G.~B.,  Teoh A.,   Hobbs M.,  2005, \mn@doi [\aj]
  {10.1086/428488}, 129, 1993

\bibitem[\protect\citeauthoryear{McClure-Griffiths, Dickey, Gaensler, Green,
  Haynes  \& Wieringa}{McClure-Griffiths et~al.}{2000}]{McClure-Griffiths2000}
McClure-Griffiths N.~M.,  Dickey J.~M.,  Gaensler B.~M.,  Green a.~J.,  Haynes
  R.~F.,   Wieringa M.~H.,  2000, \mn@doi [\aj] {10.1086/301413}, 119, 21

\bibitem[\protect\citeauthoryear{McClure-Griffiths, Dickey, Gaensler  \&
  Green}{McClure-Griffiths et~al.}{2001}]{McClureGriffiths2001}
McClure-Griffiths N.~M.,  Dickey J.~M.,  Gaensler B.~M.,   Green A.~J.,  2001,
  \mn@doi [\apj] {10.1086/323859}, 562, 424

\bibitem[\protect\citeauthoryear{McClure-Griffiths, Dickey, Gaensler  \&
  Green}{McClure-Griffiths et~al.}{2002}]{McClure-Griffiths2002}
McClure-Griffiths N.~M.,  Dickey J.~M.,  Gaensler B.~M.,   Green A.~J.,  2002,
  \mn@doi [\apj] {10.1086/342470}, 578, 176

\bibitem[\protect\citeauthoryear{McClure-Griffiths, Ford, Pisano, Gibson,
  Staveley-Smith, Calabretta, Dedes  \& Kalberla}{McClure-Griffiths
  et~al.}{2006}]{Mcclure-Griffiths2006}
McClure-Griffiths N.~M.,  Ford A.,  Pisano D.~J.,  Gibson B.~K.,
  Staveley-Smith L.,  Calabretta M.~R.,  Dedes L.,   Kalberla P. M.~W.,  2006,
  \mn@doi [\apj] {10.1086/498706}, 638, 196

\bibitem[\protect\citeauthoryear{McClure-Griffiths et~al.,}{McClure-Griffiths
  et~al.}{2009}]{McClure-Griffiths2009}
McClure-Griffiths N.~M.,  et~al., 2009, \mn@doi [\apjs]
  {10.1088/0067-0049/181/2/398}, 181, 398

\bibitem[\protect\citeauthoryear{McClure-Griffiths, Madsen, Gaensler, McConnell
   \& Schnitzeler}{McClure-Griffiths et~al.}{2010}]{McClure-Griffiths2010}
McClure-Griffiths N.~M.,  Madsen G.~J.,  Gaensler B.~M.,  McConnell D.,
  Schnitzeler D. H. F.~M.,  2010, \mn@doi [\apj] {10.1088/0004-637X/725/1/275},
  725, 275

\bibitem[\protect\citeauthoryear{Moss, Mcclure-Griffiths, Braun, Hill  \&
  Madsen}{Moss et~al.}{2012}]{Moss2012}
Moss V.~A.,  Mcclure-Griffiths N.~M.,  Braun R.,  Hill A.~S.,   Madsen G.~J.,
  2012, \mn@doi [\mnras] {10.1111/j.1365-2966.2012.20538.x}, 421, 3159

\bibitem[\protect\citeauthoryear{Neuh{\"{a}}user \& Forbrich}{Neuh{\"{a}}user
  \& Forbrich}{2008}]{Neuhauser2008}
Neuh{\"{a}}user R.,  Forbrich J.,  2008, Bo Reipurth

\bibitem[\protect\citeauthoryear{Norman \& Ikeuchi}{Norman \&
  Ikeuchi}{1989}]{Norman1989}
Norman C.~A.,  Ikeuchi S.,  1989, \mn@doi [\apj] {10.1086/167912}, 345, 372

\bibitem[\protect\citeauthoryear{Ntormousi, Dawson, Hennebelle  \&
  Fierlinger}{Ntormousi et~al.}{2017}]{Ntormousi2017}
Ntormousi E.,  Dawson J.~R.,  Hennebelle P.,   Fierlinger K.,  2017, \mn@doi
  [\aap] {10.1051/0004-6361/201629268}, 599, A94

\bibitem[\protect\citeauthoryear{Oppermann et~al.,}{Oppermann
  et~al.}{2012}]{Oppermann2012}
Oppermann N.,  et~al., 2012, \mn@doi [\aap] {10.1051/0004-6361/201118526}, 542,
  A93

\bibitem[\protect\citeauthoryear{Oppermann et~al.,}{Oppermann
  et~al.}{2015}]{Oppermann2015}
Oppermann N.,  et~al., 2015, \mn@doi [\aap] {10.1051/0004-6361/201423995}, 575,
  A118

\bibitem[\protect\citeauthoryear{Pidopryhora, Lockman  \& Shields}{Pidopryhora
  et~al.}{2007}]{Pidopryhora2007}
Pidopryhora Y.,  Lockman F.~J.,   Shields J.~C.,  2007, \mn@doi [\apj]
  {10.1086/510521}, 656, 928

\bibitem[\protect\citeauthoryear{Purcell et~al.,}{Purcell
  et~al.}{2015}]{Purcell2015}
Purcell C.~R.,  et~al., 2015, \mn@doi [\apj] {10.1088/0004-637X/804/1/22}, 804,
  22

\bibitem[\protect\citeauthoryear{Reynolds, Tufte, Haffner, Jaehnig  \&
  Percival}{Reynolds et~al.}{1998}]{Reynolds1998}
Reynolds R.~J.,  Tufte S.~L.,  Haffner L.~M.,  Jaehnig K.,   Percival J.~W.,
  1998, \mn@doi [\pasa] {10.1071/AS98014}, 15, 14

\bibitem[\protect\citeauthoryear{Schlegel, Finkbeiner  \& Davis}{Schlegel
  et~al.}{1998}]{Schlegel1998}
Schlegel D.~J.,  Finkbeiner D.~P.,   Davis M.,  1998, \mn@doi [\apj]
  {10.1086/305772}, 500, 525

\bibitem[\protect\citeauthoryear{Shull, Jones, Danforth  \& Collins}{Shull
  et~al.}{2009}]{Shull2009}
Shull J.,  Jones J.~R.,  Danforth C.~W.,   Collins J.~A.,  2009, \mn@doi [\apj]
  {10.1088/0004-637X/699/1/754}, 699, 754

\bibitem[\protect\citeauthoryear{Slavin \& Cox}{Slavin \&
  Cox}{1992}]{Slavin1992}
Slavin J.~D.,  Cox D.~P.,  1992, \mn@doi [\apj] {10.1086/171412}, 392, 131

\bibitem[\protect\citeauthoryear{Sokoloff, Bykov, Shukurov, Berkhuijsen, Beck
  \& Poezd}{Sokoloff et~al.}{1998}]{Sokoloff1998}
Sokoloff D.~D.,  Bykov a.~a.,  Shukurov A.,  Berkhuijsen E.~M.,  Beck R.,
  Poezd a.~D.,  1998, \mn@doi [\mnras] {10.1046/j.1365-8711.1999.02161.x}, 299,
  189

\bibitem[\protect\citeauthoryear{Stil, Wityk, Ouyed  \& Taylor}{Stil
  et~al.}{2009}]{Stil2009}
Stil J.,  Wityk N.,  Ouyed R.,   Taylor A.~R.,  2009, \mn@doi [\apj]
  {10.1088/0004-637X/701/1/330}, 701, 330

\bibitem[\protect\citeauthoryear{Su, Slatyer  \& Finkbeiner}{Su
  et~al.}{2010}]{Su2010}
Su M.,  Slatyer T.~R.,   Finkbeiner D.~P.,  2010, \mn@doi [\apj]
  {10.1088/0004-637X/724/2/1044}, 724, 1044

\bibitem[\protect\citeauthoryear{Suad, Caiafa, Arnal  \& Cichowolski}{Suad
  et~al.}{2014}]{Suad2014}
Suad L.~A.,  Caiafa C.~F.,  Arnal E.~M.,   Cichowolski S.,  2014, \mn@doi
  [\aap] {10.1051/0004-6361/201323147}, 564, A116

\bibitem[\protect\citeauthoryear{Sun, Han, Reich, Reich, Shi, Wielebinski  \&
  F{\"{u}}rst}{Sun et~al.}{2007}]{Sun2007a}
Sun X.~H.,  Han J.~L.,  Reich W.,  Reich P.,  Shi W.~B.,  Wielebinski R.,
  F{\"{u}}rst E.,  2007, \mn@doi [\aap] {10.1051/0004-6361:20066001}, 463, 993

\bibitem[\protect\citeauthoryear{Sun, Reich, Han, Reich, Wielebinski, Wang  \&
  M{\"{u}}ller}{Sun et~al.}{2011}]{Sun2011}
Sun X.~H.,  Reich W.,  Han J.~L.,  Reich P.,  Wielebinski R.,  Wang C.,
  M{\"{u}}ller P.,  2011, \mn@doi [\aap] {10.1051/0004-6361/201015383}, 527,
  A74

\bibitem[\protect\citeauthoryear{Tomisaka}{Tomisaka}{1990}]{Tomisaka1990}
Tomisaka K.,  1990, \mn@doi [\apj] {10.1086/185814}, 361, L5

\bibitem[\protect\citeauthoryear{Tomisaka}{Tomisaka}{1998}]{Tomisaka1998}
Tomisaka K.,  1998, \mn@doi [\mnras] {10.1046/j.1365-8711.1998.01654.x}, 298,
  797

\bibitem[\protect\citeauthoryear{Turtle, Pugh, Kenderdine  \&
  Pauliny-Toth}{Turtle et~al.}{1962}]{Turtle1962}
Turtle A.~J.,  Pugh J.~F.,  Kenderdine S.,   Pauliny-Toth I. I.~K.,  1962,
  \mn@doi [\mnras] {10.1093/mnras/124.4.297}, 124, 297

\bibitem[\protect\citeauthoryear{Uyaniker, F{\"{u}}rst, Reich, Reich  \&
  Wielebinski}{Uyaniker et~al.}{1999}]{Uyaniker1999}
Uyaniker B.,  F{\"{u}}rst E.,  Reich W.,  Reich P.,   Wielebinski R.,  1999,
  \mn@doi [\aaps] {10.1051/aas:1999494}, 138, 31

\bibitem[\protect\citeauthoryear{Valls-Gabaud}{Valls-Gabaud}{1998}]{Valls-Gabaud1998}
Valls-Gabaud D.,  1998, \mn@doi [\pasa] {10.1071/AS98111}, 15, 111

\bibitem[\protect\citeauthoryear{Wolleben \& Reich}{Wolleben \&
  Reich}{2004}]{Wolleben2004}
Wolleben M.,  Reich W.,  2004, \mn@doi [\aap] {10.1051/0004-6361:20040561},
  427, 537

\makeatother
\end{thebibliography}




\appendix

\section{Appendix}
\subsection{T-T Plot}\label{sec:tt}
\begin{figure}
	\centering
	\includegraphics[width=\columnwidth]{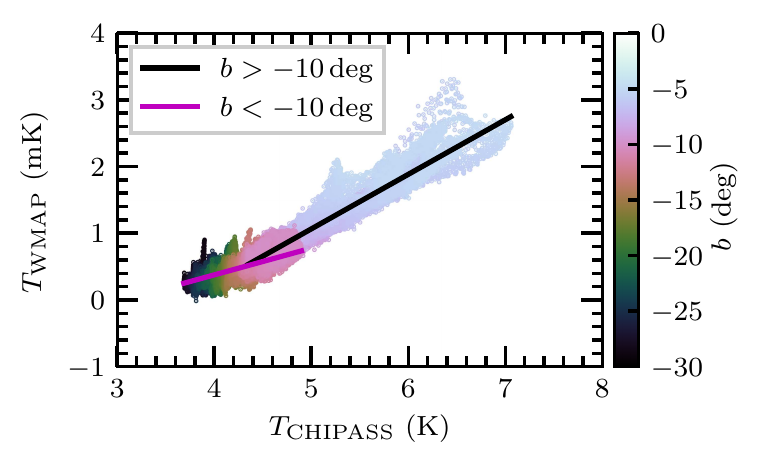}
	\caption{T-T plot: 1.4\,GHz (CHIPASS) against 23\,GHz (WMAP) total intensity (Stokes $I$) in the region of \bub. Linear fit to $b>-10\deg$ (black): $m=8.2\times10^{-4}$, $\beta=-2.6$, $R^2=0.9$. Linear fit to $b<-10\deg$ (magenta): $m=4.0\times10^{-4}$, $\beta=-2.8$, $R^2=0.5$. Note: point source at $l,b\approx[9\deg,-19\deg]$ is excluded. Here $R^2$ is the statistical coefficient of determination. Here we find that the slope of the T-T plot steepens towards the Galactic plane; which is due to non-thermal emission from warm gas in the disc. This emission, as well as emission from point sources, also causes `filaments' in the T-T plot and increases the overall scatter.}
	\label{fig:ttplot}
\end{figure}

There is an increase in the slope of the T-T plot above $b\approx-10\deg$, as shown in Figure~\ref{fig:ttplot}, which occurs above $T_\text{CHIPASS}\approx5\,$K. Additionally, there are multiple `bifurcations' present in the scatter,  As such, the slope, as fitted to this T-T plot, does not correspond to the synchrotron emission alone. Additionally, there is significant noise present below $b\approx-10\deg$ in the WMAP data; resulting in a poor fit with a coefficient of determination $R^2=0.5$. 

\subsection{Derivation of path length through shell}\label{sec:path}
To obtain the path length $L$ through \bub\ we consider a spherical shell with an inner radius of $R_i$, a thickness of $R_s$, and therefore an outer radius of $R_o = R_i+R_s$. We consider both the near and far side of the shell. We also consider the centre of the shell to be at a distance of $d_c$ from the Sun and at a Galactic latitude of $b=b_c$ and longitude $l=l_c$. Initially, we calculate $L$ through the line through the centre of shell at $l=l_c$, and then use axisymmetry to find $L$ over the entire shell. Taking $l=l_c$ we obtain the following equation:

\begin{equation}
R^2=(r\cos{b}-d_c\cos{b_c})^2+(r\sin{b}-d_c\sin{b_c})^2
\label{eqn:radiigeneral}
\end{equation}
Where $R=R_o=R_i+R_s$ for the outer boundary, and $R=R_i$ for the inner boundary. This simplifies to:
\begin{equation}
R^2=r^2-2d_cr[\cos{(b-b_c)}]+d_c^2
\label{eqn:radiigeneralsimp}
\end{equation}
Now, solving for $r$
\begin{equation}
r=d_c\cos{(b-b_c)}\pm\sqrt{\left(R^2-d_c^2\sin^2{(b-b_c)}\right)}
\label{eqn:raduisobs}
\end{equation}
The two exact solutions to this equation correspond to the near ($ r_{-} $) and far ($ r_{+} $) intersections with the line of sight. Such solutions only exist within the considered boundary (outer boundary if $R=R_o$ and inner boundary if $R = R_i$). So, the path-length through the shell ($L$) is the chord between these two boundaries and is given by:
\begin{equation}
	L(l_c,b) = \left( r_{+}(R_o)-r_{+}(R_i)\right) + \left( r_{-}(R_i) - r_{-}(R_o)\right) 
	\label{eqn:chordin}
\end{equation}
when the line of sight intersects both boundaries. When the line of sight intersects only the outer boundary the path length is:
\begin{equation}
L(l_c,b) = r_{+}(R_o)-r_{-}(R_o)
\label{eqn:chordout}
\end{equation}
To obtain the path-length as a function of $l$ and $b$ ($L(l,b)$), we assume spherical symmetry. Meaning we simply apply our solution for $L$ on the line $l=l_c$ axisymmetrically across the entire region. 

\begin{figure}
	\centering
	\includegraphics[width=\columnwidth]{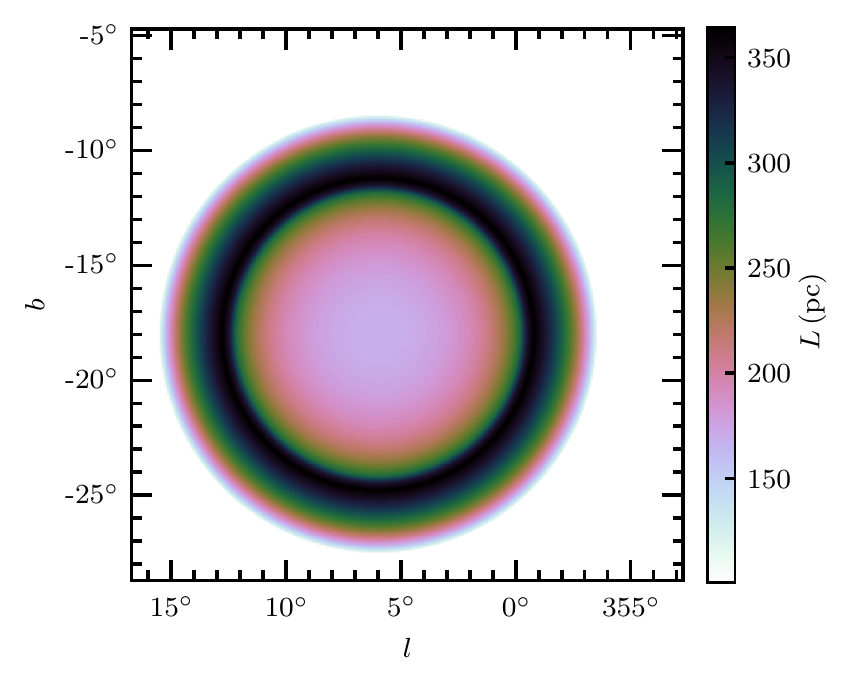}
	\caption{A map of the path-length ($L$) through \bub, as modelled by Equation~\ref{eqn:chordin} and \ref{eqn:chordout}. Here we adopt a distance to the shell of $d_c=1.5\,$kpc. The inner and outer radii ($R_i, R_o$) were chosen to match the inner outer bounds of the shell as projected on the sky. These have radii of $6.4\deg$ and $9.6\deg$, respectively.}
	\label{fig:shell}
\end{figure}

\subsection{Emission Measure map}
\begin{figure}
	\centering
	\includegraphics[width=\columnwidth]{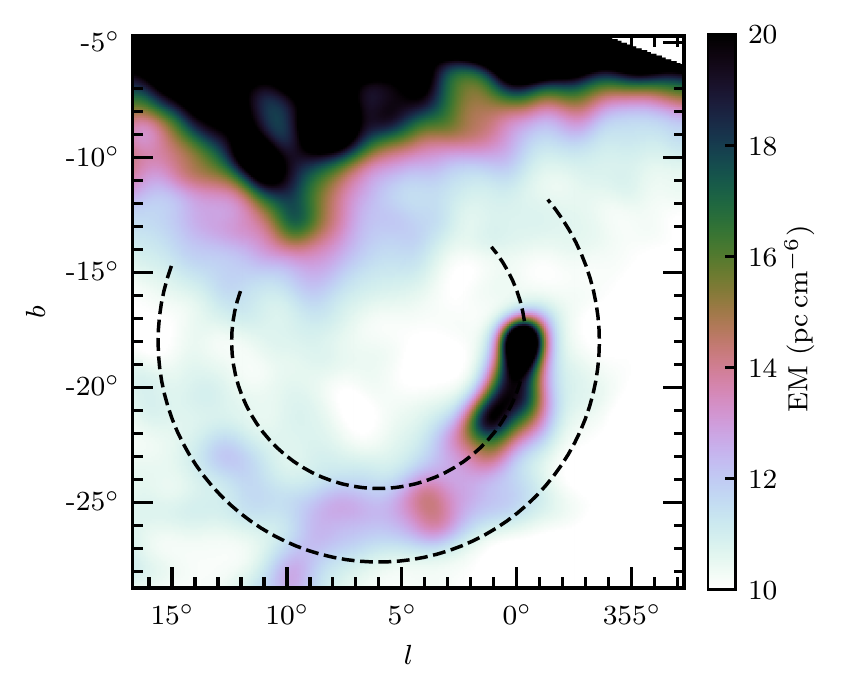}
	\caption{Map of emission measure (EM) in the region of \bub. The black dashed lines give the inner and outer bounds of the shell.}
	\label{fig:em}
\end{figure}

Figure~\ref{fig:em} shows the map of EM as obtained using Equation~\ref{eqn:emhalpha}. This map has been smoothed to the spatial resolution of WMAP and then further smoothed with a $1\deg$ Gaussian to match the other data used.


\bsp	
\label{lastpage}
\end{document}